\documentclass[preprint2]{aastex}\newcommand{\emul}{false}

\newcommand{\IncludeFigures}{true}

\usepackage{amsmath}
\usepackage{amssymb}
\usepackage{xspace}
\usepackage{lscape}

\usepackage{ifthen}
\newboolean{emul}%
\setboolean{emul}{\emul}%

\ifthenelse{\boolean{emul}}{
  \usepackage[longnamesfirst]{natbib}
  \bibpunct[authoryear]{(}{)}{;}{a}{}{,}
  \usepackage{emulateapj}
  \usepackage{apjfonts}
  \setcounter{secnumdepth}{5}
  \usepackage{graphicx}
  \submitted{Draft \today}
}

\lefthead{HEGER \& WOOSLEY}
\righthead{NUCLEOSYNTHESIS IN POP III}

\newcommand{\g}{{\ensuremath{\mathrm{g}}}\xspace}
\newcommand{\K}{{\ensuremath{\mathrm{K}}}\xspace}
\newcommand{\cm}{{\ensuremath{\mathrm{cm}}}\xspace}
\newcommand{\yr}{{\ensuremath{\mathrm{yr}}}\xspace}
\newcommand{\km}{{\ensuremath{\mathrm{km}}}\xspace}
\newcommand{\Msun}{{\ensuremath{\mathrm{M}_{\odot}}}\xspace}
\newcommand{\Sec}{{\ensuremath{\mathrm{s}}}\xspace}
\newcommand{\erg}{{\ensuremath{\mathrm{erg}}}\xspace}
\newcommand{\ergs}{{\ensuremath{\erg\,\Sec^{-1}}}\xspace}

\newcommand{\kms}{{\ensuremath{\km\,\Sec^{-1}}}\xspace}

\newcommand{\gcc}{{\ensuremath{\g\,\cm^{-3}}}\xspace}

\newcommand{\lSect}[1]{{\label{sec:#1}}}
\newcommand{\lFig}[1]{{\label{fig:#1}}}
\newcommand{\lEq}[1]{{\label{eq:#1}}}
\newcommand{\lTab}[1]{{\label{tab:#1}}}
\newcommand{\pFig}[1]{{\placefigure{fig:#1}}}

\newcommand{\Tabff}[1]{{\ref{tab:#1}}}
\newcommand{\Tab}[1]{{Table~\Tabff{#1}}}
\newcommand{\Tabs}[1]{{Tables~\Tabff{#1}}}
\newcommand{\pan}[1]{{\textit{#1}}}

\newcommand{\FIGFF}[2]{{\ref{fig:#2}\pan{#1}}}
\newcommand{\Figff}[1]{{\FIGFF{}{#1}}}
\newcommand{\FIG}[2]{{Fig.~\FIGFF{#1}{#2}}}
\newcommand{\Fig}[1]{{\FIG{}{#1}}}
\newcommand{\FIGS}[2]{{Figs.~\FIGFF{#1}{#2}}}
\newcommand{\Figs}[1]{{\FIGS{}{#1}}}

\newcommand{\Sectff}[1]{{\ref{sec:#1}}}
\newcommand{\Sect}[1]{{Sect.~\Sectff{#1}}}

\newcommand{\I}[2]{{\isotope{}{#1}{#2}}}

\newcommand{\Ep}[1]{{\ensuremath{10^{#1}}}}
\newcommand{\E}[1]{{\ensuremath{\powersep\Ep{#1}}}}

\newcommand{\Tc}{{\ensuremath{T_{\mathrm{c}}}}\xspace}
\newcommand{\Dc}{{\ensuremath{\rho_{\mathrm{c}}}}\xspace}
\newcommand{\Ec}{{\ensuremath{\eta_{\mathrm{c}}}}\xspace}

\newcommand{\Ye}{{\ensuremath{Y_e}}\xspace}

\newcommand{\MHe}{{\ensuremath{M_{\mathrm{He}}}}\xspace}

\newcommand{\MZAMS}{{\ensuremath{M_{\mathrm{ZAMS}}}}\xspace}

\def\ltaprx {\lower .1ex\hbox{\rlap{\raise .6ex\hbox{\hskip .3ex
        {\ifmmode{\scriptscriptstyle <}\else 
                {$\scriptscriptstyle <$}\fi}}}
        \kern -.4ex{\ifmmode{\scriptscriptstyle \sim}\else 
                {$\scriptscriptstyle\sim$}\fi}}}
\def\gtaprx {\lower .1ex\hbox{\rlap{\raise .6ex\hbox{\hskip .3ex
        {\ifmmode{\scriptscriptstyle >}\else 
                {$\scriptscriptstyle >$}\fi}}}
        \kern -.4ex{\ifmmode{\scriptscriptstyle \sim}\else 
                {$\scriptscriptstyle\sim$}\fi}}}

\ifthenelse{\boolean{emul}}{\renewcommand{\pFig}[1]{}}

\newcommand{\FigyeFile}{f1.eps} \newcommand{\Figye} {Yields of the
dominant elements (\textsl{left scale}) and explosion energies
(\textsl{thick gray line, right scale}; one ``foe'' is \Ep{51}\,\erg,
about the explosion energy of a typical modern supernova) as a
function of helium core mass \citep[see also][]{heg01}.  The range
shown corresponds to main sequence masses of $\sim140-260\,\Msun$.
Helium cores of lower mass do not explode in a single pulse and those
of higher mass collapse into black holes (see \Fig{MM}).  \lFig{ye}}

\newcommand{\FigMMFile}{f2.eps} \newcommand{\FigMM} {Initial-final
mass function (IFMF) of non-rotating primordial stars ($Z=0$).  The
\textsl{x-axis} gives the initial stellar mass.  The \textsl{y-axis}
gives both the final mass of the collapsed remnant (\textsl{thick
black curve}) and the mass of the star when the event begins that
produces that remnant (e.g., mass loss in AGB stars, supernova
explosion for those stars that make a neutron star, etc.;
\textsl{thick gray curve}).  We distinguish four regimes of initial
mass: \emph{low mass stars} below $\sim10\,\Msun$ that end as white
dwarfs; \emph{massive stars} between $\sim10\,\Msun$ and
$\sim100\,\Msun$; \emph{very massive stars} between $\sim100\,\Msun$
and $\sim1000\,\Msun$; and \emph{supermassive stars} (arbitrarily)
above $\sim1000\,\Msun$.  Since no mass loss is expected for $Z=0$
stars before the final stage, the grey curve is approximately the same
as the line of no mass loss (\textsl{dotted}).  Exceptions are
$\sim100-140\,\Msun$ where the pulsational pair-instability ejects the
outer layers of the star before it collapses, and above
$\sim500\,\Msun$ where pulsational instabilities in red supergiants
may lead to significant mass loss \citep{BHW01}.  Since the magnitude
of the latter is uncertain, lines are \textsl{dashed}.  In the low
mass regime we assume, even in Z = 0 stars, that mass loss on the
asymptotic giant branch (AGB) leads to the star losing its envelope
and becoming a CO or NeO white dwarf (though the mechanism and thus
the resulting initial-final mass function may differ from solar
composition stars).  ``Massive stars'' are defined as stars that
ignite carbon and oxygen burning non-degenerately and do not leave
white dwarfs.  The hydrogen-rich envelope and parts of the helium core
(\textsl{dash-double-dotted curve}) are ejected in a supernova
explosion.  Below an initial mass of $\sim25\,\Msun$, neutron stars
are formed. Above that, black holes form, either in a delayed manner
by fall back of the ejecta, or directly during iron core collapse
(above $\sim40\,\Msun$).  The defining characteristic of {\sl very}
massive stars is the electron-positron \emph{pair instability} after
carbon burning.  This begins as a pulsational instability for helium
cores of $\sim 40$\,\Msun ($\MZAMS\sim100$\,\Msun).  As the mass
increases, the pulsations become more violent, ejecting any remaining
hydrogen envelope and an increasing fraction of the helium core
itself. An iron core can still eventually form in hydrostatic
equilibrium in such stars, but it collapses to a black hole. Above
$\MHe = 63$\,\Msun or about $\MZAMS = 140\,\Msun$, and on up to $\MHe
= 133$\,\Msun or about $\MZAMS = 260$\,\Msun) a single pulse disrupts
the star.  Above 260\,\Msun, the pair instability in non-rotating
stars results in complete collapse to a black hole.  \lFig{MM}}

\newcommand{\FigpfFile}{f3.eps}
\newcommand{\Figpf}{Production factors for very massive stars (helium
cores of 65--130\,\Msun, corresponding to initial masses of
$\sim140-260\,\Msun$) integrated over an IMF and compared to solar
abundances as a function of element number, $Z$.  The
integration assumed a Salpeter-like IMF with three different
exponents: -0.5 (\textsl{thick end of triangle}), -1.5 (\textsl{solid
dot}), and -3.5 (\textsl{thin end of triangle}).  \lFig{pf}}

\newcommand{\FigpfmFile}{f4.eps}
\newcommand{\Figpfm}{Production factors for massive stars
(12--40\,\Msun; \textsl{dotted line and open triangles}) integrated
over IMF and compared with solar abundances as a function of element
number.  The yields are taken from \citet{WW95} and in this plot we
use the low explosion energy primordial models of the ``A'' series,
Z12A, Z15A, $\ldots$\,.  The \textsl{solid line and filled triangles})
give the same integration but also including exploding very massive
stars ($\sim140-260\,\Msun$).  In the mass range 40--100\,\Msun 
essentially the whole helium core falls into a black hole,
ejecting only the unprocessed envelope.  In the mass range
100--140\,\Msun some of the outer layers of the helium core may be
ejected, adding to the carbon and oxygen yields and maybe a
little to the neon and magnesium yields, but not to the
heavier elements.  The IMF is assumed Salpeter-like with an exponent
of -1.5.  \lFig{pfm}}

\newcommand{\FigpfMFile}{f5.eps}
\newcommand{\FigpfM}{Same as figure \Fig{pfm}, but here we used the
high-energetic explosion of the ``B'' series from \cite{WW95} for
25\,\Msun and above, Z25B, Z30B, $\ldots$\,.  
\lFig{pfM}}

\slugcomment{Submitted to {\em The Astrophysical Journal}}

\begin{document}

\title{The Nucleosynthetic Signature of Population III}
\author{A.\ Heger and S.\ E.\ Woosley}

\vskip 0.2 in
\affil{Department of Astronomy and Astrophysics \\
University of California, Santa Cruz, CA 95064}
\authoremail{alex@ucolick.org, woosley@ucolick.org}

\begin{abstract} 

Growing evidence suggests that the first generation of stars may have
been quite massive ($\sim100-300\,\Msun$).  Could these stars have
left a distinct nucleosynthetic signature?  We explore the
nucleosynthesis of helium cores in the mass range $\MHe=64$ to
133\,\Msun, corresponding to main-sequence star masses of
approximately 140 to 260\,\Msun.  Above $\MHe=133\,\Msun$, without
rotation and using current reaction rates, a black hole is formed and
no nucleosynthesis is ejected.  For lighter helium core masses,
$\sim$40 to 63\,\Msun, violent pulsations occur, induced by the pair
instability and accompanied by supernova-like mass ejection, but the
star eventually produces a large iron core in hydrostatic equilibrium.
It is likely that this core, too, collapses to a black hole, thus
cleanly separating the heavy element nucleosynthesis of pair
instability supernovae from those of other masses, both above and
below.  Indeed, black hole formation is a likely outcome for all
Population III stars with main sequence masses between about 25 \Msun
and 140\,\Msun ($M_{\rm He}$ = 9 to 63\,\Msun) as well as those above
260\,\Msun. Nucleosynthesis in pair-instability supernovae varies
greatly with the mass of the helium core which determines the maximum
temperature reached during the bounce.  At the upper range of
exploding core masses, a maximum of 57\,\Msun of $^{56}$Ni is produced
making these the most energetic and the brightest thermonuclear
explosions in the universe.  Integrating over a distribution of
masses, we find that pair instability supernovae produce a roughly
solar distribution of nuclei having even nuclear charge (Si, S, Ar,
etc.), but are remarkably deficient in producing elements with odd
nuclear charge - Na, Al, P, V, Mn, etc.  This is because there is no
stage of stable post-helium burning to set the neutron excess. Also,
essentially no elements heavier than zinc are produced owing to a lack
of $s$- and $r$-processes.  The Fe/Si ratio is quite sensitive to
whether the upper bound on the IMF is over 260\,\Msun or somewhere
between 140 and 260\,\Msun.  When the yields of pair-instability
supernovae are combined with reasonable estimates of the
nucleosynthesis of Population III stars from 12 to 40\,\Msun, this
distinctive pattern of deficient production of odd-Z elements
persists. Some possible strategies for testing our predictions are
discussed.

\end{abstract}


\section{Introduction}
\lSect{intro}

Simulations of the collapse of primordial molecular clouds suggest
that the first generation of stars \citep{OG96} contained many massive
members, from one hundred up to one thousand solar masses
\citep[e.g.,][]{Lar99,BCL99,ABN00}.  Calculations by \citet{NU00}
suggest a bi-modal initial mass function for Population III with peaks
at $\sim$100\,\Msun and 1 - 2\,\Msun.  Considerable attention has been
given recently to the nucleosynthesis of stars having main sequence
mass below 100\,\Msun \citep{WW95,TNH96}, but what is the
nucleosynthetic signature of stars appreciably heavier?

For non-rotating stars over 260\,\Msun on the main sequence, the
answer is simple - zero \citep{FWH01}.  The nuclear energy released
when the star collapses on the pair instability is not sufficient to
reverse the implosion before the onset of the
photodisintegration-instability and the star becomes a black hole
\citep[see also][]{RSZ67,BAC84,GFE85,Woo86}, sweeping all heavy
element production inside.
%
Between approximately 140 and 260\,\Msun lies the domain of
pair-instability supernovae.  After central helium burning, stars have
high enough central entropy that they enter a temperature and density
regime in which electron/positron pairs are created in abundance,
converting internal gas energy into rest mass of the pairs without
contributing much to the pressure \citet{BRS67,BAC84}.  When this
instability is encountered, the star contracts rapidly until implosive
oxygen and silicon burning, depending on the mass of the star, produce
enough energy to revert the collapse.  These objects then completely
disrupt in nuclear-powered explosions.
%
Unlike their lighter cousins, the explosion mechanism in
pair-instability supernovae is well understood and there are no issues
of ``mass-cut'' or ``fall-back'' to complicate the outcome.  The
stellar core implodes to a certain maximum temperature that depends on
its mass, burns fuel explosively due to inertial overshoot, and then
explodes.  Provided such stars existed and retained their high mass
until death, the outcome (neglecting rotation) is uniquely calculable.
It is thought that the pre-collapse winds and pulsations of such
stars result in little mass loss \citep{Kud00,BHW01,VKL01}.

Nucleosynthesis in pair-instability supernovae has been previously
studied by \citet{OEF83,WW82,Woo86}, but those calculations examined a
limited range of stellar masses and had very restricted nuclear
reaction networks. In particular, the synthesis of rare elements with
odd nuclear charge was not followed, nor were species heavier than
nickel.

Here we follow the evolution of 14 helium stars in the mass range 65
to 130\,\Msun using a nuclear reaction networks of 304 (presupernova)
and 477 (supernova) isotopes. The synthesis of isotopes at the
boundaries of the networks is always sufficiently low to determine the
yields of these objects completely.  We also briefly describe the
evolution of helium cores that are slightly larger (over 133.3\,\Msun)
and slightly smaller (63\,\Msun) with the conclusion that black hole
formation is likely in both cases. This allows one to separate cleanly
the yields of pair-instability supernovae from stars of slightly lower
or higher mass.  In the following sections we discuss our computational
procedure and the nucleosynthetic results.  In \Sect{results} these
results are integrated over an initial mass function for comparison
with observed abundances in very metal-deficient stars and clouds. The
distinctive signature is a composition almost solar-like in the ratios
among elements with even nuclear charge, but lacking in elements with
odd charge.  We also discuss the possibility that the IMF is
truncated somewhere short of $\sim$200\,\Msun in which case the
contribution to iron group nucleosynthesis from pair-instability
supernovae would be relatively small.

\pFig{MM}
\ifthenelse{\boolean{emul}}{
\vspace{1.5\baselineskip}
\noindent
\includegraphics[angle=0,width=\columnwidth]{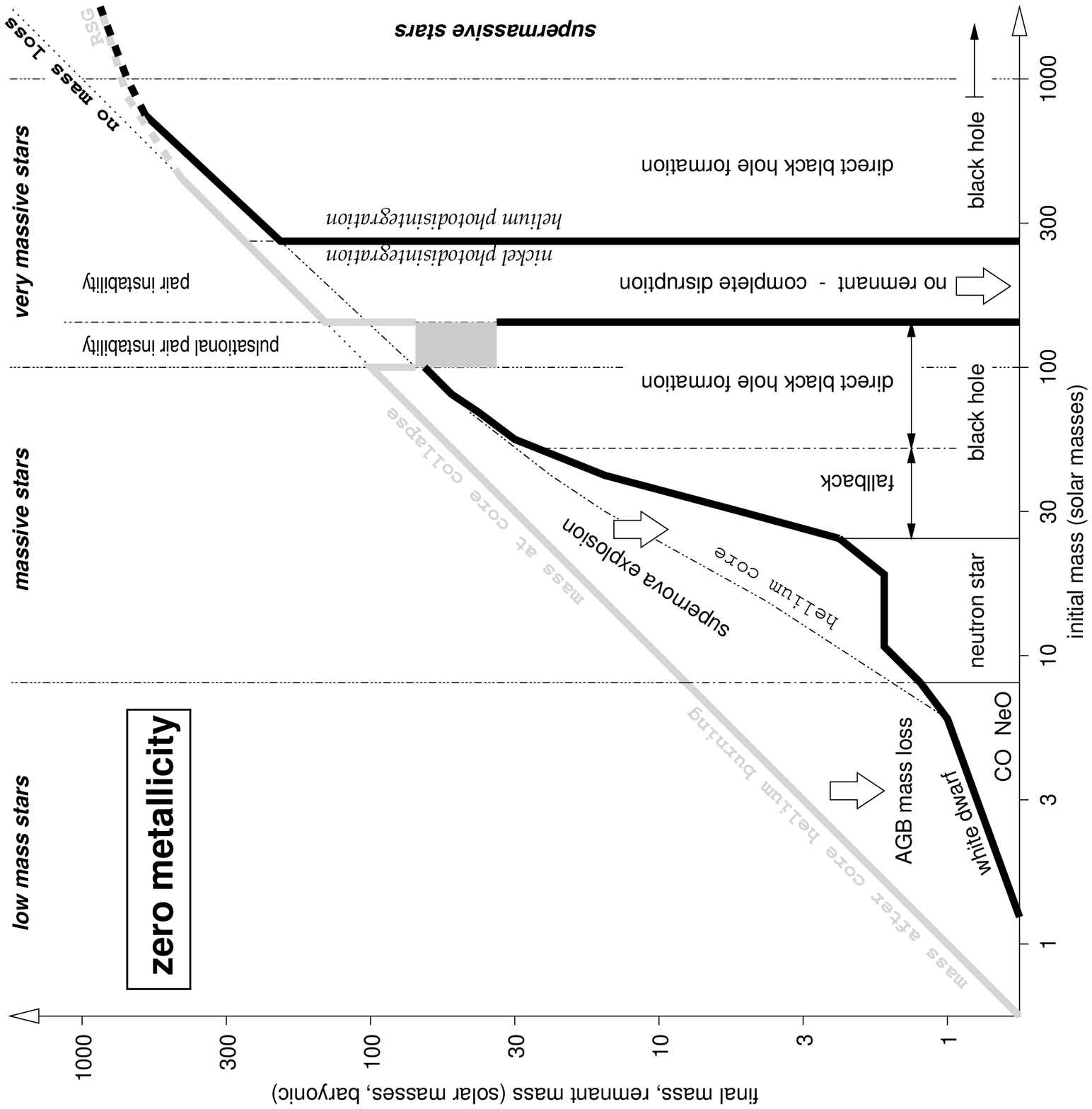}
\figcaption{\FigMM}
\vspace{1.5\baselineskip}}{}

\section{Computational Procedure and Assumptions}

All calculations were carried out using the implicit hydrodynamics
code, KEPLER \citep{WZW78,HLW00}.  The equation of state allows for
electrons and pairs of arbitrary degeneracy and relativity.  The
opacity was taken from \citet{IR96} and the principal nuclear reaction
rates were as described by \citet{rau01}.  For the important
$^{12}$C($\alpha,\gamma)^{16}$O reaction rate, we used \citet{Buc96}
updated with Buchmann (2000, priv. com; see also \citealp{CF88}, and
\citealp{kun01}).  Mass loss was not taken into account, except what
was driven by the pair instability pulsations studied here.

Nucleosynthesis was determined by ``co-pro\-cess\-ing'' the stellar
evolution model with a network of 304 isotopes prior to the explosion
and 477 nuclei during in the explosion. The network contained all the
necessary isotopes of elements from hydrogen to ruthenium.  Trial
calculations with a larger network showed that this network was
adequate to correctly follow all species that accumulated to greater
than 10$^{-15}$ by mass fraction throughout the presupernova evolution
and the explosion. 

A series of helium core masses was then calculated, starting from the
helium burning main sequence and ending when either: \textbf{1)} an
iron core formed in hydrostatic equilibrium and began to collapse on
the photo-disintegration instability; \textbf{2)} a black hole was
clearly about to form; or \textbf{3)} the star was completely
disrupted by an explosion.  Our helium stars were presumed to consist
initially of almost pure helium (Pop III).  A total of 25 helium cores
were studied having masses 60, 63, 64, 65, 70, 75, 80 ... 125, 130,
131, 132, 133, 133.3, 133.31, 134, 135, 140\,\Msun.  This spans the
range of objects expected to become pair instability supernovae and
explores the boundary of this range with good resolution.  The initial
composition (\Tab{iabu}) used in all cases was that of a 200\,\Msun
Pop III star (93\,\Msun helium core) evolved from the hydrogen burning
main sequence to helium ignition.  Some CNO isotopes formed as a
consequence of $3 \alpha$ reactions on the main sequence
\citep{EC71,BHW01}.  The neutron excess of this starting composition,
$\eta = \Sigma (N_i -Z_i)Y_i$, was 1.9\E{-7}.  A typical value of this
parameter for solar metallicity stars at the onset of helium burning
is 0.002.  As we shall see this change has dramatic consequences for
the nucleosynthesis of odd-Z elements.

\section{Stellar Models and Yields}

Following helium burning, each stellar core consisted mostly of
oxygen. For example, the mass fractions of $^{12}$C and $^{16}$O when
the star reached a central temperature of $5 \times 10^8$ K were
[0.10, 0.82], [0.089, 0.81], and [0.080, 0.80] for helium stars of 80,
100, and 120 \,\Msun, respectively.  The remainder was mostly
\I{20}{Ne} and \I{24}{Mg}.  The total oxygen masses in each of these
three stars at this same point was 61.9, 77.0, and 91.6 \, \Msun. The
surface of each star remained nearly pure helium, but the helium mass
fraction declined to 0.5 within 1.0 \, \Msun of the surface.  Had mass
loss been included these would have become WC or WO stars.

Each star then proceeded to burn away its central carbon abundance -
and then neon - radiatively.  That is, no exoergic, convective core
was formed.  By the end of carbon burning the star had begun to
encounter the pair instability, and by central neon depletion,
typically at (2.0--2.2)\E9\,\K, its carbon/oxygen core was collapsing
at speeds in excess of 1000\,\kms.  Explosion followed, powered by
explosive oxygen burning, and, for higher masses, additionally by
explosive silicon burning.  Bounce temperatures and densities for the
80, 100, and 120\,\Msun models were 3.88, 4.53, and 5.39 \E9\,\K and
2.32, 3.20, and 5.42 \E6\,\gcc, respectively.  \Tab{dense} gives the
central temperature, \Tc, density, \Dc, and the neutron, proton, and
\I4{He} mass fractions at the time of highest central density.
\Fig{ye} gives the explosion energy (kinetic energy at infinity) and
bulk nucleosynthetic yields for all stars studied.  Greater
nucleosynthetic detail can be found in \Tab{y} and is discussed in
\Sect{results}.

\pFig{ye}
\ifthenelse{\boolean{emul}}{
\vspace{1.5\baselineskip}
\noindent
\includegraphics[angle=0,width=\columnwidth]{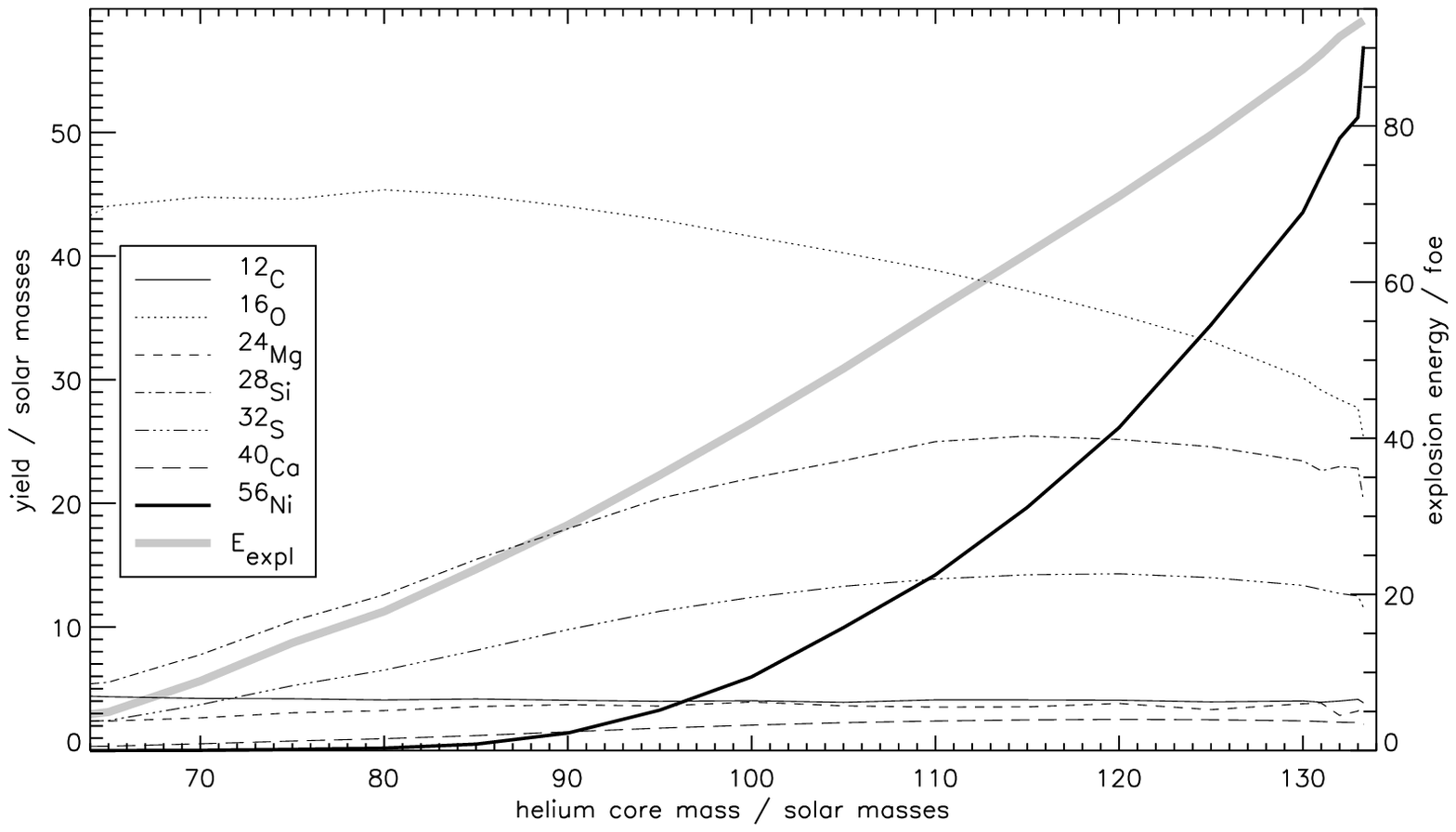}
\figcaption{\Figye}
\vspace{1.5\baselineskip}}{}

\subsection{Critical Masses - Stars at the Extreme}
\lSect{critmass}

Although the answer depends upon the rate for
\I{12}C($\alpha$,$\gamma$)\I{16}O, the neglect of rotation, and
possibly the convection model, we determined the range of helium core
masses that become pair instability supernovae to be $\sim$64 to
$\sim$133\,\Msun.  This corresponds to main sequence stars of
approximately 140--260\,\Msun.  An approximate empirical relation
between the helium core mass and main sequence mass in this mass range
which we use here is
\begin{equation}
\MHe \approx \frac{13}{24} \left(\MZAMS - 20\,\Msun\right)
\lEq{MMHe}
\;.
\end{equation}

The 63\,\Msun helium core had an interesting evolution that included
three violent episodes of mass ejection spanning an interval of five
thousand years, before finally settling down to burn silicon in
hydrostatic equilibrium. The first collapse happened, as in other
pair-unstable stars, just after helium burning. The center of the star
reached a temperature of 3.2\E9\,\K and density 1.5\E6\,\gcc before
rebounding in an energetic explosion that released 6.5\E{51}\,\erg
from nuclear burning, principally explosive oxygen burning
(4.5\E{51}\,\erg was released above a central temperature of 3\E9\,\K;
6.4\E{51}\,\erg was released above 2\E9\,\K; 6.5\E{51}\,\erg is the
net energy release after nuclear burning first exceeded neutrino
losses in the star).  The total binding of the star actually became
briefly positive at this point (6.9\E{50}\,\erg), but the outward
motion was hydrodynamically concentrated into the outer layers with
the result that only 12.8\,\Msun was ejected with kinetic energy
1.2\E{51}\,\erg, the equivalent of an ordinary Type II supernova.  The
remaining star then experienced an extended period of Kelvin-Helmholtz
contraction during which no fuel was burned, but the central
temperature and density gradually rose.  This contraction phase after
the first ``pulse'' was slow because temperature and density at
the stellar center after expansion ($\rho=25$\,\gcc; $T=7.3\E7$\,\K)
were too low for neutrino emission.  Thus the star contracted on the
Kelvin-Helmholtz time-scale for photon emission from the surface and
was unaffected by neutrino emission.  During most of the next
4800\,\yr the star had a surface luminosity of 8--10 \E{39}\,\ergs and
that luminosity governed the evolution until late times.  However, the
details of the re-expansion and consequently the time-scale of the
subsequent contraction phase may depend on the modeling of energy
transport (convection/Rayleigh-Taylor instabilities) during the
dynamic phases of the core.  We assumed convection in subsonic
regions.

Forty-eight hundred years after the first pulse, a second collapse and
explosion occurred, this time ejecting 2.7\,\Msun with kinetic energy
of only 1.3\E{50}\,\erg.  The star did not expand so much following
this weak explosion, rebounding from a temperature of 3.68\E9\,\K,
$\rho = 1.0$\E7\,\gcc to only 1.2\E9\,\K and 4.9\E5\,\gcc.  Neutrino
losses rapidly robbed the core of energy so that the remaining star
(now 47.5\,\Msun) evolved rapidly through a second Kelvin-Helmholtz
phase.  By this point the central 4\,\Msun of the star was already
composed of more than 50\% oxygen burning products, chiefly silicon
and sulfur, and the mass fraction of iron in the stellar center was
0.1 (\I{54}{Fe}).

Eight days later, a third and final violent pulse occurred, powered
again by off-center explosive oxygen burning. This pulse went to a
central temperature of 4.5\E9\,\K and density 2.0\E6\,\gcc and was
more violent than the second (but less than the first).  It ejected
2.2\,\Msun with kinetic energy 5.1\E{50}\,\erg.  The expansion also
was more extreme, central conditions declining to the point
(8.5\E8\,\K; 1.9\E5\,\gcc) where neutrino losses were not very
efficient (though still greater than the surface luminosity by orders
of magnitude). At this point the oxygen depleted core was 5\,\Msun and
the inner 0.8\,\Msun of the star was mostly iron.

The final contraction phase followed lasted 1.8 years. The stellar
core then settled into stable silicon shell burning with no more
pair-instability pulses. Silicon burned in a series of shells, each
encompassing about 0.2\,\Msun (zoning was 0.01\,\Msun) until the iron
core grew to 1.8\,\Msun and collapsed on the photodisintegration
instability.  It should be noted, however, that this collapsing core
differed in a variety of ways from those found in lower mass stars:
\textbf{1)} there was no steep decline in density around the iron
core, nor sharp increase in entropy usually associated with the oxygen
burning shell. The oxygen shell was in fact at 5.3\,\Msun; \textbf{2)}
the entropy in the core and its surrounding were anomalously large
ranging from 1 at the center to 3 at the edge of the iron core and 6.0
at 3\,\Msun; \textbf{3)} the central electron mole number, $\Ye =
0.446$ was unusually large and the value in the outer core, larger
still, 0.490; and \textbf{4)} the net binding energy outside the iron
core was $-3.9 \times 10^{51}$ erg. All these differences go in the
direction of a less centrally condensed structure that will be harder
to explode \citep{wil86,Fry99}. We believe that, in the absence of
very appreciable rotation, this star will become a black hole.

A similar behavior was observed for the 45\,\Msun helium core of a
100\,\Msun main sequence star studied by \citet{Woo86}.  That star
experienced 4 pulses resulting from the pair instability before
settling down to burn silicon stably and also become a black hole
\citep{wil86}.  The pulses were less violent though with a total
duration of only 1.6\,\yr. This probably marks the lower extremity of
the pulsational pair domain, but stars in between might have
intermediate energies and durations.  The relevance of such objects
for e.g., SN 1961V \citep[e.g.,][]{fil95} and Eta-Carina \citep{dav99}
remains largely unexplored.  Lacking radioactivity, the ejecta might
not be bright, but this would be greatly altered if the explosion
occurred inside a star that still had an extended hydrogen envelope or
when the ejected shells ran into one another.

On the upper boundary of the mass range for pair instability, violent
explosions are found that leave no bound remnant.  A 133.3\,\Msun
helium core exploded with a kinetic energy of 93.5\E{51}\,\erg and
ejected 57.0\,\Msun of \I{56}{Ni}. This is approximately 100 times the
$^{56}$Ni produced by a typical Type Ia supernova and, for a bare
helium star, might have a peak luminosity 100 times as great (i.e.,
$\sim\Ep{45}$\,\ergs).  Even inside a blue supergiant, as one might
expect for Pop III stars at death, the radioactive display would be
spectacular, brighter than a galaxy. The explosion would be the most
energetic thermonuclear explosion in the universe.  At the point of
highest density (1.9\E7\,\gcc) in this model, a central temperature of
7.4\E9\,\K is reached and the star contains 19.4\,\Msun of \I4{He},
0.98\,\Msun of protons, and 0.054\,\Msun of neutrons from
photodisintegration, corresponding to about 5\E{52}\,\erg that is
released when they recombine to \I{56}{Ni}. During the collapse phase
the helium-rich layer collapses rapidly onto the helium-free core and
helium is burned implosively even up to silicon as the dominant
product.

However, going just a small step up in mass, for a non-rotating star,
gives no explosion at all. For a 133.31\,\Msun helium star the infall
did not turn around into a explosion, but instead continued after a
brief phase of slowing, into a black hole. With rotation, the
evolution of all these stars could be greatly altered
\citep{GFE85,SW88,FWH01}.

Immediately below the \emph{lower} boundary of the mass range for
pair-instability supernovae, it is also likely that black holes are
the main product.  This is because either a successful outgoing shock
fails to be launched for stars above about 40\,\Msun \citep{Fry99} or
the shock is so weak that fall-back converts the neutron star into a
black hole. Either way the star fails to eject most of its heavy
elements.  \citet{WW95} find, for an assumed constant explosion energy
of about 1.2\E{51}\,\erg, that the final collapsed remnant in Pop III
stars of masses 22, 25, 30, 35, and 40\,\Msun are 1.5, 6.4, 8.2, 13,
and 17\,\Msun respectively. Apparently above about 30 \Msun the heavy
element synthesis of ordinary Pop III supernovae (i.e., not
pair-unstable) is negligible.

The fact that the pair-supernova domain is bounded above and below by
stars that make black holes (\Fig{MM}) and the fact that the explosion
is easy to calculate makes the nucleosynthesis from this mass range
easy to determine unambiguously provided one specifies an IMF.

\subsection{Nucleosynthetic Results}
\lSect{results}

The yields for all the models are given in \Tab{y} for each stable
nucleus created in any appreciable abundance compared with the
sun. Values are given in solar masses after all unstable progenitors
have decayed. The value given for \I{56}{Fe}, in particular, is
actually that of \I{56}{Ni}.  The reaction network included sufficient
weak interactions to follow accurately the evolving neutron
excess. Final central values for the central value of $\eta$ for our
fiducial helium stars of 80, 100, and 120\,\Msun are 0.00034, 0.0016
and 0.0037.  Even for 130\,\Msun the value was 0.0052.  As one moves
out from the center the neutron excess declines rapidly.  These small
values ensure that \I{56}{Fe} and other abundant nuclei like
\I{48}{Ti}, \I{52}{Cr}, and \I{60}{Ni} are made as their $Z = N$
progenitors.  In the last column of \Tab{dense} the central neutron
excess, \Ec, is given at the time of maximum density of the star.
 
\Tab{pf} gives the production factors corresponding to these
yields. The production factor, $P_i$, is defined as the ratio of the
mass fraction of a given species in the ejecta compared to its mass
fraction in the sun. One implication is that ``solar production'' of a
given isotope could occur in a star having metallicity say $[Z] = -3$
(0.1\,\% that of the sun) if a fraction, \Ep{-3}/$P_i$, of that star's
mass had passed through pair-instability supernovae of the given type.
Of course one must normalize to the greatest $P_i$ and all species
having smaller $P_i$ would be underproduced relative to the sun.

Already obvious in \Tabs{y} and \Tabff{pf} is the small production of
nuclei that require excess neutrons for their existence. This includes
all nuclei with odd charge above \I{14}N: \I{23}{Na}, \I{27}{Al},
\I{31}P, and the like - as well as neutron-rich isotopes like
\I{29,30}{Si}, \I{33,34,36}S, \I{38}{Ar}, etc.  Such nuclei all
require a neutron excess for their production. They are underproduced
here because of the low value of $\eta$ in all regions of all stars
except the deep interiors of the most massive explosions (thus
\I{54}{Fe} has an appreciable yield in the 130\,\Msun model). Also
absent is any appreciable nucleosynthesis for $A \gtaprx 66$.

Elements above the iron group are absent because there was no
\textsl{s}- or \textsl{r}-process in our stars.  The
\textsl{r}-process requires very rapid expansion time scales from
extremely high temperature, high entropy, and a large neutron
excess. None of these are realized here. The \textsl{s}-process
requires a neutron source - such as $^{22}$Ne($\alpha$,n)\I{25}{Mg} or
\I{13}C($\alpha$,n)\I{16}O, and heavy element ``seed nuclei''. There
are no seed nuclei in our Pop III stars and very little \I{13}C or
\I{22}{Ne} is present during helium burning. [Since we studied helium
stars, any possible production of \I{13}C that might result from the
inter-penetration of hydrogen envelope and convective helium shell was
not followed. This has never been demonstrated to give neutrons in a
massive star, but might be worth further investigation].

The neutron excess is low in our stars because: 1) The assumed Pop III
starting composition (\Tab{iabu}) implies a very small value of $\eta$
at the end of helium burning.  This $\eta$ comes about from the
conversion of \I{14}N into \I{18}O early in helium burning by
\I{14}N($\alpha$,$\gamma$)\I{18}F(e$^+$,$\nu$)\I{18}O. The abundance of
CNO is very low in Pop III stars); and 2) subsequent burning stages
occur too rapidly and at too low a density for much additional
electron capture or positron decay \citep{AT69,TA71,Arn73}.

This neutron deficiency imprints a distinctive signature on the
nucleosynthesis of Pop III pair-instability supernovae that is even
more extreme than seen in metal free stars of lower mass.

\subsection{Integrated Nucleosynthesis}

In \Fig{pf}, the individual isotopes in \Tab{iabu} have been summed,
divided by the mass ejected (equal to the mass of the star in the
present study), and the resultant elemental mass fraction integrated
over an estimated Salpeter-like \citep{Sal59} initial mass function
(IMF) for the assumed progenitor stars of the helium cores
(\Sect{critmass}) and divided by their solar mass fraction.  We show
the result of this integration for three different slopes,
$\gamma=-0.5$, -1.5, and -3.5 of the IMF, where $\gamma$ is defined by
the number of stars formed per mass interval,
$\gamma\equiv 1+\mathrm{d}\log N/\mathrm{d}\log M$.  The ``dot''
indicates the middle value, connected by lines, and the ``thick'' and
``thin'' ends of the triangle show the shallower and steeper IMF
slopes, respectively.

\pFig{pf}
\ifthenelse{\boolean{emul}}{
\vspace{1.5\baselineskip}
\noindent
\includegraphics[angle=0,width=\columnwidth]{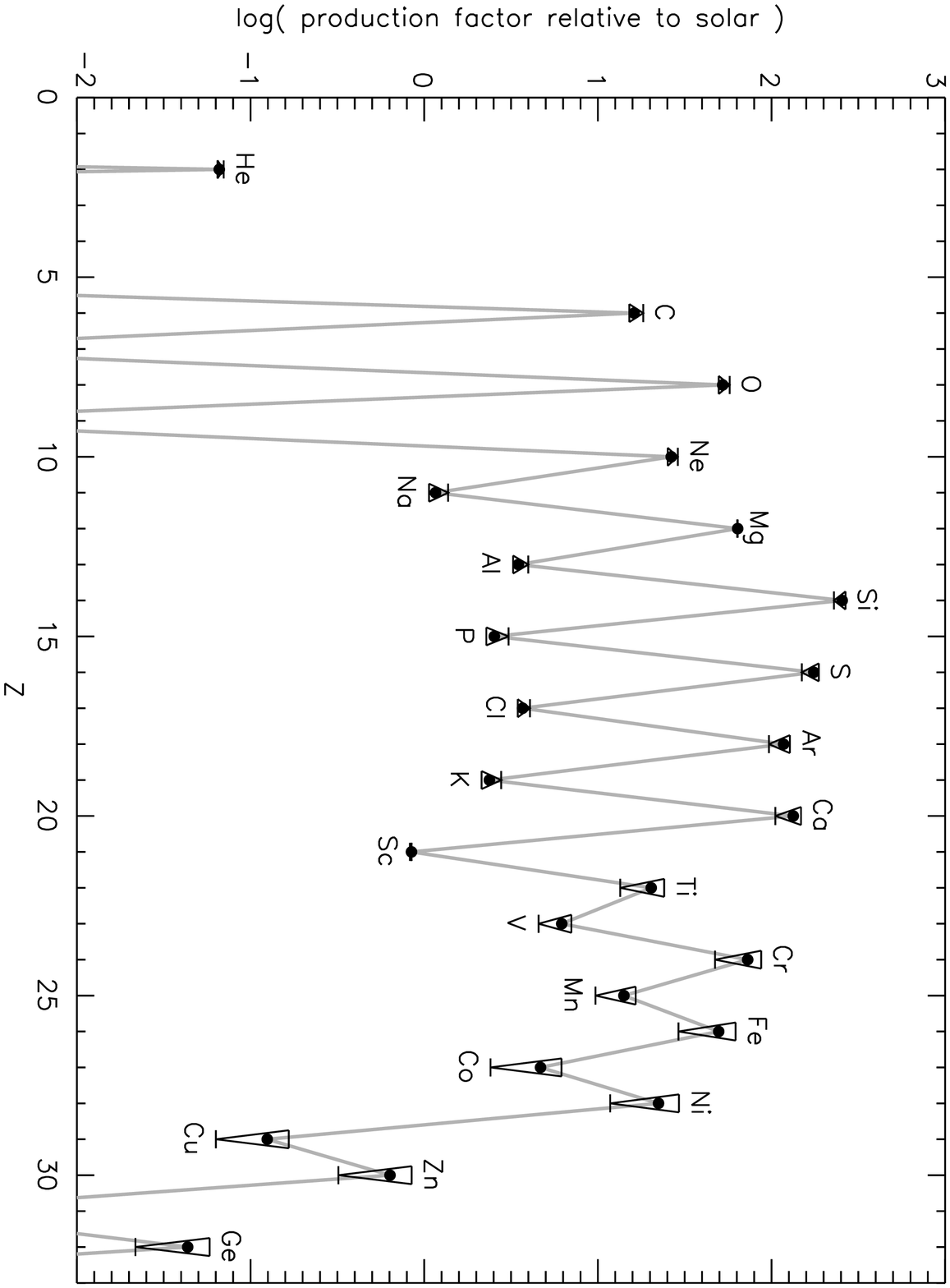}
\figcaption{\Figpf}
\vspace{1.5\baselineskip}}{}

The results, shown in \Fig{pf}, are not very sensitive to the slope of
the IMF since we are only studying stars in a limited mass range (a
factor of two from the lowest to the highest mass considered).  The
dependence of the integrated production factor on the element number,
$Z$, shows the odd-even effect discussed in \Sect{results} and
quantifies it to be from one order of magnitude for iron group
elements to two orders of magnitude for some of the intermediate mass
elements.  The iron group shows a smaller effect because of the weak
interactions that occurs in the central regions of the more massive
cores which reach high density during their bounce.

Given the unusual nature of the synthesis site (pair-instability
supernovae are not generally thought to be the dominant site where
solar abundances were produced), the overall approximate agreement the
yields with the solar abundances of elements with even charge is
somewhat surprising.  The nuclear properties of the elements - their
binding energies and cross sections - are apparently as important as
the stellar environment. However there are differences. For example,
Si and S are overabundant compared to O, Fe, and Mg.  As noted in
\Sect{results}, elements above Ni are essentially absent.  Above Ge,
the numbers are all below the lower bound of the plot and their
abundance decreases exponentially with mass number.

This conflicts with observations which show appreciable
\textsl{r}-process elements present even in very metal-deficient stars
\citep[e.g.,][]{bur00}, at least for [Fe/H] $\gtaprx$ -2.9.  Since we
believe that the \textsl{r}-process requires neutron stars - one way
or another - for its production, the synthesis of elements by lower
mass stars must also be considered.  That is, no matter what the IMF
for Pop III, abundances at [Fe/H] $\gtaprx$ -3 could not have been
made solely by pair-instability supernovae.

Also, lacking any hard evidence of what the IMF for Pop III
was like, one cannot preclude a truncation somewhere within the mass
range $\MHe=64$\ldots133\,\Msun.  Taking away the stars above
$\MHe=90\,\Msun$ could clearly give an arbitrarily large production
factor for oxygen and intermediate mass elements compared to the iron
group.

\pFig{pfm}
\ifthenelse{\boolean{emul}}{
\vspace{1.5\baselineskip}
\noindent
\includegraphics[angle=0,width=\columnwidth]{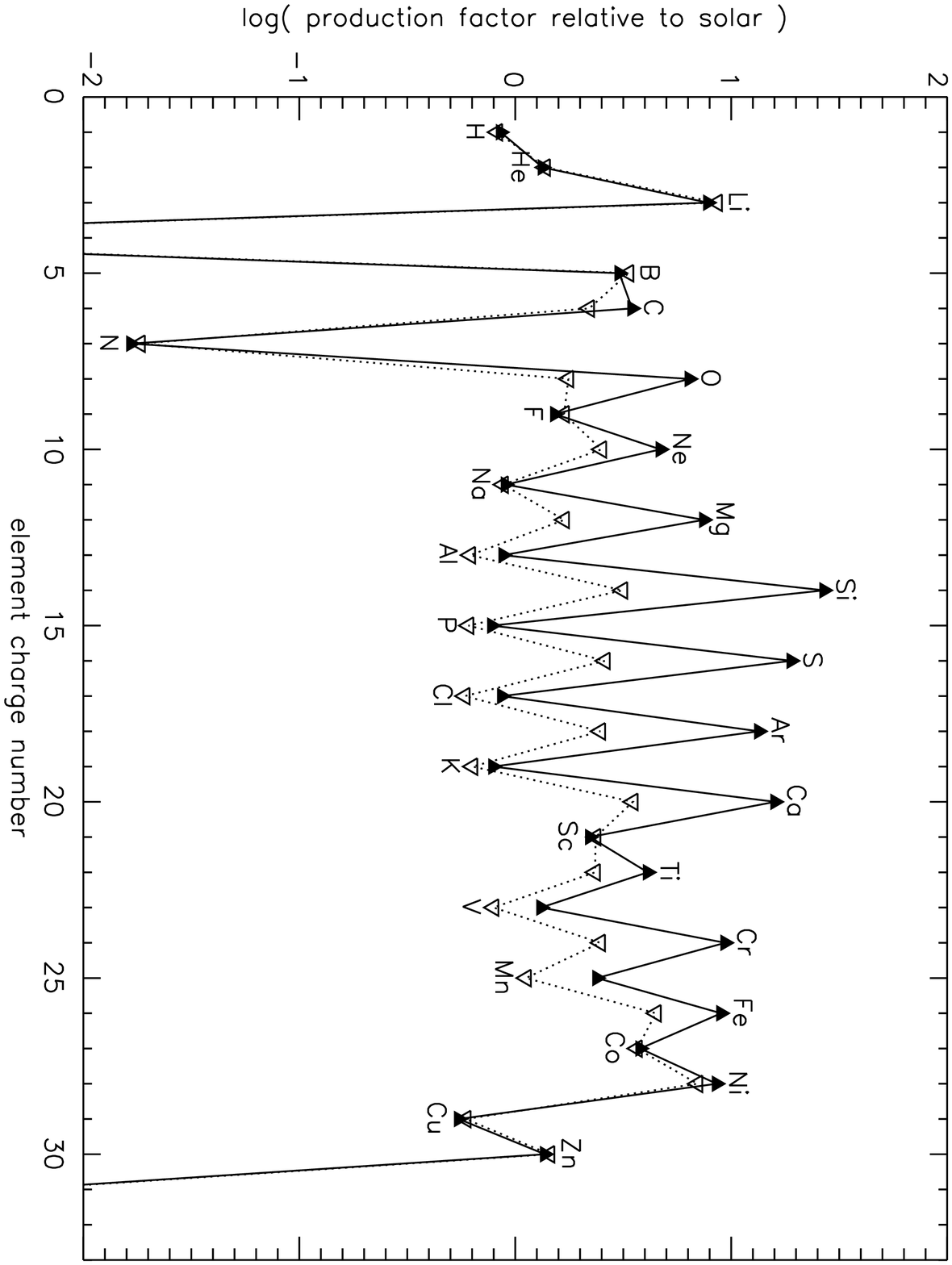}
\figcaption{\FigpfM}
\vspace{1.5\baselineskip}}{}

To explore further the consequences of an admixture of lighter Pop III
stars, we included in our integration the yields of the ``Z-series''
of Pop III supernovae studied by \citet{WW95}, metal-free stars in the
mass range 12 to 40\,\Msun.  \Figs{pfm} and \Figff{pfM} show the
consequences of including these stars with two different choices of
explosion energy. For larger explosion energies, less fall back occurs
and more iron group elements are ejected.  In the higher mass stars
which tend to make black holes, more explosion energy ejects more
heavy elements of all kinds.  This explains the difference between
\Figs{pfm} and \Figff{pfM}.  We also assume, as the calculations
suggest, that for reasonable supernova energies, no heavy elements are
ejected in the explosion of Pop III stars over 40\,\Msun
\citep{Fry99}.  This is in part because very little mass loss occurs
and the star dies while still having a large helium core.

The figures show, not too surprisingly, that including the ``lower
mass'' supernovae tends to wash out the dramatic odd-even effect seen
in the pair-instability supernovae.  Still the two mass ranges, $\MZAMS
= $12--40 and 140--260, make approximately the same total masses of
heavy elements. The lighter stars are more abundant in a Salpeter IMF,
but they eject less mass - both because they are themselves less
massive and because of fall back.  The modification introduced by a
pair-unstable massive component may have observational consequences so
long as the high mass component continues to form.  Of course the
formation of pair unstable objects may cease after the creation of
only a small metallicity \citep{BHW01}.

\pFig{pfM}
\ifthenelse{\boolean{emul}}{
\vspace{1.5\baselineskip}
\noindent
\includegraphics[angle=0,width=\columnwidth]{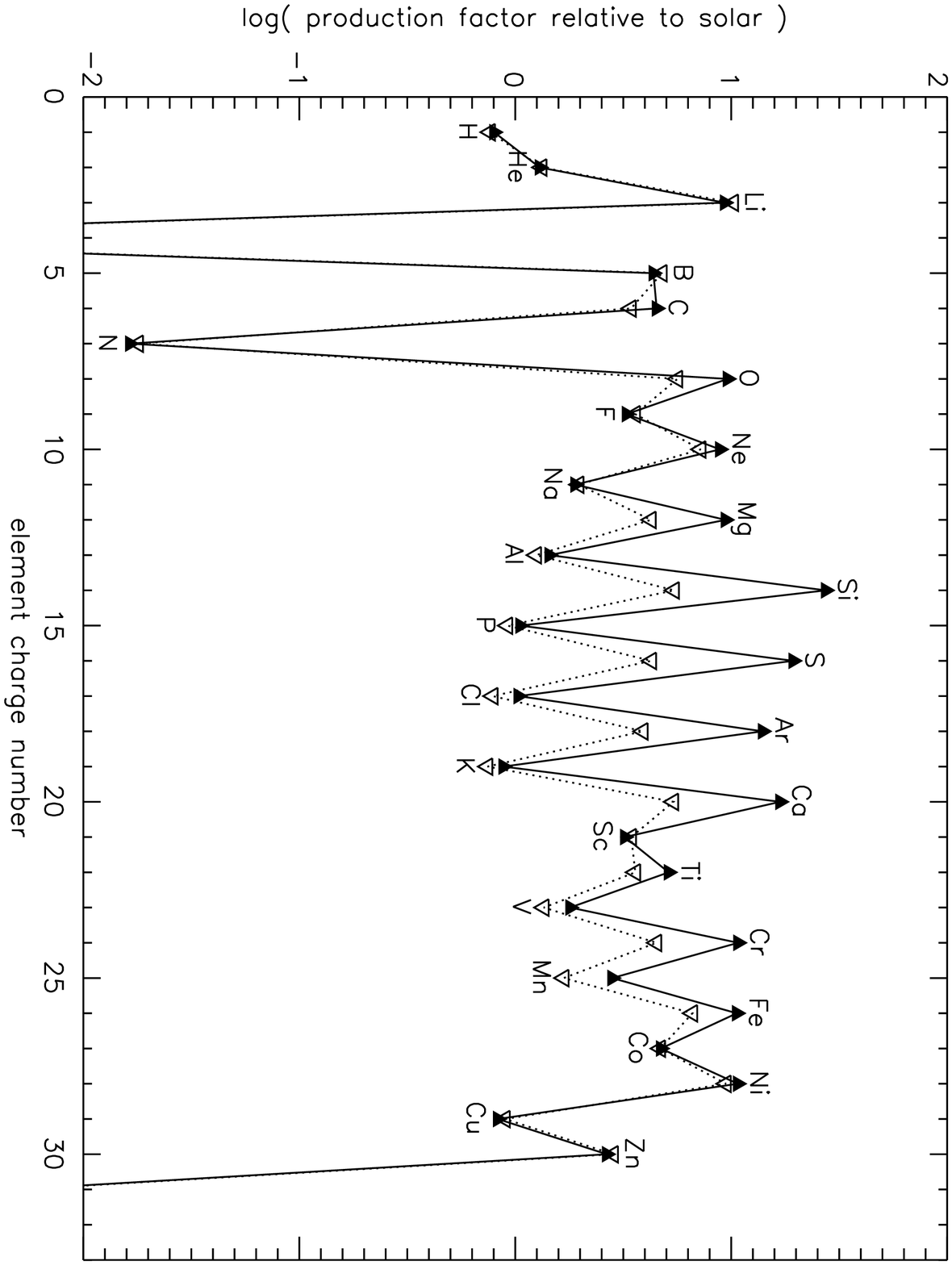}
\figcaption{\Figpf}
\vspace{1.5\baselineskip}}{}

\subsection{Some Specific Nuclei}

\subsubsection{Li, B, F}

These fragile light nuclei are made only (in our calculations) by the
neutrino-process in stars lighter than about 40\,\Msun.  They are not
produced in pair-instability supernovae

\subsubsection{Nitrogen}
\lSect{nitro}

An important issue in very metal-deficient stars is the production of
primary nitrogen \citep[see, e.g.,][]{NRB01}.  In this paper we have
studied only helium cores and thus will have missed any nitrogen
production that occurs when the helium convective shell and hydrogen
envelope of these same massive stars mix \citep{WW82}.  We hope to
return to this matter in a future publication, but note for now the
sensitivity of the results to uncertain parameters of convective
overshoot and rotationally-induced mixing \citep{WW95,HWW00,UNN00}.
It may also be possible to produce primary nitrogen in AGB stars
\citep{ven01,DW01}.

\subsubsection{Zinc}

Very little zinc is produced in the pair-instability explosions. What
is made is produced chiefly as \I{64,66}{Zn} made as \I{64,66}{Ge} in
an extreme alpha-rich freeze out at the upper end of the mass range
that explodes. However, one does expect zinc production in the
neutrino-powered wind just after the shock is launched in the deepest
layers of stars between about 10 and 20\,\Msun \citep{hof96},
essentially the same site as the $r$-process.  This neutrino wind was
not included either in the present calculation or that of
\citet{WW95}, but could easily give a much higher prediction for zinc.

\subsubsection{The $r$- and $s$-processes} 

Because of the lack of any heavy seed nuclei or any appreciable
neutron source in helium burning, there is no $s$-process in low
metallicity pair-instability supernovae.  Also, because no neutron
star is formed and no shock wave passes through any region where
appreciable electron capture has occurred, there is no
$r$-process.  This latter conclusion might possibly be altered in
rapidly rotating pair-unstable stars that produce black holes,
accretion disks, and jets.

However, the simplest explanation of $r$- process abundances in
metal-deficient stars is that they indicate the contribution of an
additional nucleosynthetic component - namely those lower mass stars
which make regular supernovae and neutron stars.

\section{Conclusions and Observational Tests} 

The natural places to look for the distinctive nucleosynthetic pattern
of pair instability supernovae is in the stellar photospheres of very
metal-deficient stars nearby and in damped Lyman-alpha systems at high
redshift. However, to truly see the signature of Pop III one must go
to very low metallicities.  It is estimated by \citet{BHW01} that the
stars responsible for pair-instability supernovae are already
significantly pulsationally unstable when the metallicity has risen to
as little as $Z\sim\Ep{-4}$ solar (the lowest non-zero metallicity
investigated by \citealp{BHW01}; the stellar structure changes
significantly between Z/Z$_{\scriptscriptstyle \odot}$ = 0 and
10$^{-4}$).

Studies of giant stars in the Milky Way are beginning to push the
frontier to [Fe/H$] = -4$ \citep{mcw95,RNB96,NRB01,gir01} and reveal
some interesting tendencies - some expected and some surprising.  One
of the most striking is the dramatic decrease of [Al/Fe] (or [Al/Mg])
at very low metallicity.  Without non-LTE corrections, the deficiency
approaches 1.5\,dex \citep{NRB01} and would strongly suggest an
pair-instability contribution (\Fig{pfm}).  However, estimated
corrections bring the Al abundance up and into better accord with what
is expected from ordinary supernovae.  The ``alpha-elements'', Mg, Si,
and Ca are clearly up by 0.3 to 0.5\,dex at low metallicity,
consistent with pair-instability supernovae, but not inconsistent with
regular supernovae either.  The fact that [Ti/Fe] is up by the same
amount is surprising in any case.  Perhaps it is not so much that
these intermediate mass elements are up, as it is Fe that is down -
due to a lack of Type Ia supernovae.  Even more intriguing is the
\emph{overproduction} of [Co/Fe] at low metallicities, which is
difficult to understand in any model.  Production of Co occurs by the
$\alpha$-rich freeze out and can be enhanced by high temperatures and
rapid expansion.  Perhaps we are seeing evidence of deeper bounces in
pair-unstable stars in which rotation was important, but then one
would expect other elements made from the $\alpha$-rich freeze out,
like Ni, also to be enhanced, and they are not.  A plot of [Mn/Fe]
vs.\ [Fe/H] shows a clear decrease from nearly solar at [Fe/H$]=-2$ to
about [Mn/Fe$]=-1$ at [Fe/H$]=-3.5$ \citep{gir01}.  This is consistent
with pair-instability supernovae, but again not inconsistent with what
core collapse supernovae alone could do (\Figs{pfm} and \Figff{pfM};
see also \citealt{nak99}).

More measurements, and more precise measurements of abundances at
[Fe/H$] \approx -4$ would help, as well as a better understanding of
the non-LTE corrections involved in the spectral analysis.
Particularly sensitive to our stellar models are the ratios of pairs
of elements separated by one proton: [Na/Mg], [Al/Mg], [Al/Si], [P/S],
[P/Si], [Cl/S], [K/Ca], and the like.  One should keep in mind though
that a single pair-instability explosion that produced 40\,\Msun of
iron could provide all the iron necessary for [Fe/H]$ = -4$ in
4\E8\,\Msun of primordial material.  Mixing probably is not so
efficient at [Fe/H]$ = -4$ that we do not sample individual events.
It is thus possible to get \emph{very} large excursions in ratios like
[Si/Fe], [O/Fe], etc., even in stars where [O/Fe] is greatly
\emph{subsolar} - see the individual entries in \Tab{y}.  The effects
shown in \Figs{pf}, \Figff{pfm}, and \Figff{pfM} may need to be
applied to a statistically large sample of stars at [Fe/H]$ = -4$.  We
understand that this may be challenging.

Abundances in damped Ly-alpha systems are also starting to be measured
for heavy elements.  \citet{lu96,PW99,PW01} have measured abundances
in damped Ly-alpha systems for redshifts $> 1.5$ and metallicity down
to about [Fe/H]$ = -2$.  There is a tendency for O, Si, S, and Ti to
be high compared to solar iron, perhaps again reflecting a deficiency
of Type Ia supernova nucleosynthesis, while Cr is down by about a
factor of two and Ni is flat \citet{PW01}.  Aluminum is deficient by
up to an order of magnitude and there is also some indication that Mn
declines with redshift (Prochaska, private communication) just as it
does in metal-deficient stars.  Some of these observations are
consistent, e.g., with the solid line in \Fig{pfm}, and might indicate
the presence of a pair-instability contribution, especially the large
Si/Al ratio and moderately high Si/Fe and S/Fe abundances.  Oxygen to
iron could be boosted if the IMF were steeper at higher masses, but a
deficiency of Cr is difficult to understand and is more consistent
with nucleosynthesis in lower mass supernovae.  What is needed, are
more measurements of odd-$Z$ elements (Al, P, and Mn might be
candidates) at still smaller metallicities, though this too will be
challenging.  Interpretation of damped Ly-alpha systems will also
continue to be complicated by the uncertain effects of dust
condensation, but have to their advantage the fact that the
observations sample large aggregations of matter.

\acknowledgements We are grateful to Chris Fryer for discussions about
\Fig{MM}.  This research has been supported by the NSF (AST
97-316569), the DOE ASCI Program (B347885) and the Alexander von
Humboldt-Stiftung (FLF-1065004).

{}

\clearpage

\newpage
\onecolumn

{

\renewcommand{\E}[1]{&{\ensuremath{(#1)}}}
\newcommand{\EE}{&}
\renewcommand{\I}[2]{{\ensuremath{^{#1}}}&{\ensuremath{\mathrm{#2}}}}
\newcommand{\NoData}{\multicolumn{2}{c}{\nodata}}
\newcommand{\PPI}{\I{\phantom{99}}{\phantom{Mm}}}
\newcommand{\PPE}{\phantom{$9.99$}&\phantom{$(-99)$}}

\begin{table}
\caption{Initial Abundances\lTab{iabu}}
\begin{tabular}{cccccc}
\hline
\hline
Species   &   X       &  Species  &    X     &  Species  &    X     \\
\hline
$^4$He    & 0.9998    & $^{12}$C  & 2.0(-04) & $^{14}$C  & 1.1(-07) \\ 
$^{14}$N  & 2.0(-08)  & $^{15}$N  & 2.5(-09) & $^{16}$O  & 3.1(-06) \\ 
$^{17}$O  & 3.0(-08)  & $^{18}$O  & 1.0(-06) & $^{19}$F  & 6.9(-09) \\ 
$^{20}$Ne & 4.2(-09)  & $^{21}$Ne & 3.6(-10) & $^{22}$Ne & 6.7(-07) \\ 
$^{23}$Na & 4.1(-10)  & $^{30}$Si & 1.3(-10) & $^{31}$P  & 3.8(-10) \\ 
$^{32}$S  & 1.2(-10)  & $^{36}$S  & 2.1(-10) & $^{38}$Ar & 1.1(-10) \\ 
\hline
\end{tabular}
\end{table}

\begin{table}
\centering
\caption{Central Paramters at Maximum Density\lTab{dense}}
\begin{tabular}{rr@{}lr@{}lr@{}lr@{}lr@{}lr@{}l}
\hline
\hline
\multicolumn{1}{c}{mass} &
\multicolumn{2}{c}{\Tc} &
\multicolumn{2}{c}{\Dc} &
\multicolumn{2}{c}{$X$(n)} &
\multicolumn{2}{c}{$X$(p)} &
\multicolumn{2}{c}{$X$($^4$He)} &
\multicolumn{2}{c}{\Ec} \\
\hline
 65 &    1.741\E{  9} &    3.158\E{  5} &    1.145\E{-20} &    9.808\E{-17} &    1.756\E{-13} &    2.905\E{- 4} \\
 70 &    3.570\E{  9} &    2.001\E{  6} &    1.412\E{-12} &    2.356\E{- 4} &    5.395\E{- 5} &    2.812\E{- 4} \\
 75 &    3.867\E{  9} &    2.544\E{  6} &    2.381\E{-11} &    9.438\E{- 4} &    3.014\E{- 4} &    2.892\E{- 4} \\
 80 &    3.876\E{  9} &    2.316\E{  6} &    2.960\E{-11} &    1.040\E{- 3} &    3.492\E{- 4} &    2.583\E{- 4} \\
 85 &    4.025\E{  9} &    2.479\E{  6} &    9.769\E{-11} &    2.370\E{- 3} &    8.908\E{- 4} &    2.400\E{- 4} \\
 90 &    4.197\E{  9} &    2.699\E{  6} &    3.761\E{-10} &    5.328\E{- 3} &    2.518\E{- 3} &    2.636\E{- 4} \\
 95 &    4.355\E{  9} &    2.902\E{  6} &    1.418\E{- 9} &    8.919\E{- 3} &    6.228\E{- 3} &    2.959\E{- 4} \\
100 &    4.533\E{  9} &    3.195\E{  6} &    6.519\E{- 9} &    1.325\E{- 2} &    1.627\E{- 2} &    3.746\E{- 4} \\
105 &    4.720\E{  9} &    3.577\E{  6} &    2.775\E{- 8} &    1.743\E{- 2} &    3.399\E{- 2} &    4.705\E{- 4} \\
110 &    4.931\E{  9} &    4.079\E{  6} &    1.158\E{- 7} &    2.144\E{- 2} &    6.029\E{- 2} &    5.643\E{- 4} \\
115 &    5.140\E{  9} &    4.637\E{  6} &    4.591\E{- 7} &    2.460\E{- 2} &    9.971\E{- 2} &    7.311\E{- 4} \\
120 &    5.390\E{  9} &    5.423\E{  6} &    2.118\E{- 6} &    2.688\E{- 2} &    1.672\E{- 1} &    9.632\E{- 4} \\
125 &    5.734\E{  9} &    6.766\E{  6} &    1.421\E{- 5} &    2.757\E{- 2} &    3.002\E{- 1} &    1.317\E{- 3} \\
130 &    6.169\E{  9} &    9.012\E{  6} &    1.312\E{- 4} &    2.356\E{- 2} &    5.312\E{- 1} &    1.867\E{- 3} \\
\hline
\end{tabular}
\end{table}

\begin{landscape}

\begin{table}
\setlength{\tabcolsep}{1ex}
\centering
\caption{Yields (in solar masses)\lTab{y}}
\scalebox{0.79}{
\begin{tabular}{r@{}lr@{}lr@{}lr@{}lr@{}lr@{}lr@{}lr@{}lr@{}lr@{}lr@{}lr@{}lr@{}lr@{}lr@{}l}
\hline
\hline
\multicolumn{2}{c}{ion} &
\multicolumn{2}{c}{ 65} &
\multicolumn{2}{c}{ 70} &
\multicolumn{2}{c}{ 75} &
\multicolumn{2}{c}{ 80} &
\multicolumn{2}{c}{ 85} &
\multicolumn{2}{c}{ 90} &
\multicolumn{2}{c}{ 95} &
\multicolumn{2}{c}{100} &
\multicolumn{2}{c}{105} &
\multicolumn{2}{c}{110} &
\multicolumn{2}{c}{115} &
\multicolumn{2}{c}{120} &
\multicolumn{2}{c}{125} &
\multicolumn{2}{c}{130} \\
\hline
\I{  4}{He} &    1.96\EE     &    1.55\EE     &    1.44\EE     &    1.46\EE     &    1.44\EE     &    1.41\EE     &    1.42\EE     &    1.47\EE     &    1.53\EE     &    1.66\EE     &    1.77\EE     &    2.00\EE     &    2.38\EE     &    2.82\EE     \\
\I{ 12}{ C} &    6.89\EE     &    4.54\EE     &    4.32\EE     &    4.33\EE     &    4.28\EE     &    4.21\EE     &    4.13\EE     &    4.01\EE     &    3.85\EE     &    3.74\EE     &    3.73\EE     &    3.71\EE     &    3.61\EE     &    3.49\EE     \\
\I{ 13}{ C} &    2.33\E{- 7} &    4.19\E{- 7} &    7.44\E{- 7} &    8.86\E{- 7} &    8.48\E{- 7} &    5.78\E{- 7} &    3.59\E{- 7} &    4.31\E{- 7} &    2.30\E{- 7} &    1.60\E{- 7} &    1.64\E{- 7} &    1.51\E{- 7} &    1.38\E{- 7} &    9.36\E{- 8} \\
\I{ 14}{ N} &    7.84\E{- 5} &    5.04\E{- 5} &    4.31\E{- 5} &    4.15\E{- 5} &    3.78\E{- 5} &    3.46\E{- 5} &    3.23\E{- 5} &    3.10\E{- 5} &    2.92\E{- 5} &    2.75\E{- 5} &    2.42\E{- 5} &    1.95\E{- 5} &    1.41\E{- 5} &    1.10\E{- 5} \\
\I{ 15}{ N} &    7.27\E{- 7} &    7.03\E{- 6} &    6.69\E{- 6} &    6.69\E{- 6} &    6.57\E{- 6} &    6.45\E{- 6} &    6.41\E{- 6} &    6.43\E{- 6} &    6.51\E{- 6} &    6.76\E{- 6} &    6.99\E{- 6} &    7.19\E{- 6} &    7.03\E{- 6} &    6.65\E{- 6} \\
\I{ 16}{ O} &    4.92\E{  1} &    4.58\E{  1} &    4.44\E{  1} &    4.68\E{  1} &    4.66\E{  1} &    4.59\E{  1} &    4.52\E{  1} &    4.39\E{  1} &    4.27\E{  1} &    4.11\E{  1} &    3.98\E{  1} &    3.81\E{  1} &    3.59\E{  1} &    3.34\E{  1} \\
\I{ 17}{ O} &    4.27\E{- 6} &    9.85\E{- 7} &    7.76\E{- 7} &    7.47\E{- 7} &    6.60\E{- 7} &    5.93\E{- 7} &    5.38\E{- 7} &    5.09\E{- 7} &    4.73\E{- 7} &    4.45\E{- 7} &    4.13\E{- 7} &    3.74\E{- 7} &    3.26\E{- 7} &    2.86\E{- 7} \\
\I{ 18}{ O} &    2.81\E{- 6} &    9.94\E{- 7} &    8.50\E{- 7} &    8.15\E{- 7} &    7.94\E{- 7} &    7.84\E{- 7} &    7.63\E{- 7} &    8.10\E{- 7} &    7.99\E{- 7} &    8.23\E{- 7} &    8.35\E{- 7} &    8.50\E{- 7} &    8.89\E{- 7} &    8.47\E{- 7} \\
\I{ 19}{ F} &    1.63\E{- 8} &    1.41\E{- 8} &    1.28\E{- 8} &    1.23\E{- 8} &    1.23\E{- 8} &    1.42\E{- 8} &    1.61\E{- 8} &    1.69\E{- 8} &    1.93\E{- 8} &    2.03\E{- 8} &    2.40\E{- 8} &    2.63\E{- 8} &    2.76\E{- 8} &    2.58\E{- 8} \\
\I{ 20}{Ne} &    4.99\EE     &    4.04\EE     &    3.89\EE     &    4.22\EE     &    4.20\EE     &    4.10\EE     &    3.98\EE     &    4.06\EE     &    3.86\EE     &    3.90\EE     &    3.84\EE     &    3.88\EE     &    3.92\EE     &    3.73\EE     \\
\I{ 21}{Ne} &    5.04\E{- 4} &    6.53\E{- 5} &    2.10\E{- 4} &    2.49\E{- 4} &    2.99\E{- 4} &    2.51\E{- 4} &    1.82\E{- 4} &    2.14\E{- 4} &    1.42\E{- 4} &    1.65\E{- 4} &    1.45\E{- 4} &    1.52\E{- 4} &    1.68\E{- 4} &    1.24\E{- 4} \\
\I{ 22}{Ne} &    1.86\E{- 4} &    2.01\E{- 5} &    1.85\E{- 5} &    1.97\E{- 5} &    2.19\E{- 5} &    2.27\E{- 5} &    2.26\E{- 5} &    2.37\E{- 5} &    2.28\E{- 5} &    2.46\E{- 5} &    2.39\E{- 5} &    2.41\E{- 5} &    2.32\E{- 5} &    2.02\E{- 5} \\
\I{ 23}{Na} &    9.13\E{- 3} &    3.14\E{- 3} &    2.95\E{- 3} &    3.02\E{- 3} &    2.96\E{- 3} &    2.88\E{- 3} &    2.82\E{- 3} &    2.76\E{- 3} &    2.69\E{- 3} &    2.66\E{- 3} &    2.68\E{- 3} &    2.74\E{- 3} &    2.81\E{- 3} &    2.78\E{- 3} \\
\I{ 24}{Mg} &    1.53\EE     &    3.02\EE     &    3.49\EE     &    3.67\EE     &    3.97\EE     &    4.24\EE     &    4.38\EE     &    4.41\EE     &    4.40\EE     &    4.31\EE     &    4.50\EE     &    4.55\EE     &    4.42\EE     &    4.38\EE     \\
\I{ 25}{Mg} &    3.78\E{- 3} &    1.08\E{- 3} &    3.07\E{- 3} &    2.23\E{- 3} &    3.36\E{- 3} &    4.15\E{- 3} &    4.70\E{- 3} &    4.27\E{- 3} &    4.57\E{- 3} &    4.55\E{- 3} &    4.04\E{- 3} &    3.76\E{- 3} &    3.39\E{- 3} &    3.56\E{- 3} \\
\I{ 26}{Mg} &    3.71\E{- 3} &    1.74\E{- 3} &    1.69\E{- 3} &    1.73\E{- 3} &    1.79\E{- 3} &    1.92\E{- 3} &    2.06\E{- 3} &    1.92\E{- 3} &    2.05\E{- 3} &    1.91\E{- 3} &    1.94\E{- 3} &    1.86\E{- 3} &    1.72\E{- 3} &    1.73\E{- 3} \\
\I{ 27}{Al} &    3.37\E{- 2} &    1.77\E{- 2} &    1.66\E{- 2} &    1.59\E{- 2} &    1.67\E{- 2} &    1.72\E{- 2} &    1.77\E{- 2} &    1.63\E{- 2} &    1.66\E{- 2} &    1.55\E{- 2} &    1.50\E{- 2} &    1.42\E{- 2} &    1.29\E{- 2} &    1.34\E{- 2} \\
\I{ 28}{Si} &    3.15\E{- 1} &    7.97\EE     &    1.22\E{  1} &    1.36\E{  1} &    1.65\E{  1} &    1.92\E{  1} &    2.14\E{  1} &    2.31\E{  1} &    2.45\E{  1} &    2.52\E{  1} &    2.57\E{  1} &    2.57\E{  1} &    2.51\E{  1} &    2.42\E{  1} \\
\I{ 29}{Si} &    1.96\E{- 3} &    2.32\E{- 2} &    2.25\E{- 2} &    2.22\E{- 2} &    2.14\E{- 2} &    2.07\E{- 2} &    2.05\E{- 2} &    1.94\E{- 2} &    1.94\E{- 2} &    1.81\E{- 2} &    1.78\E{- 2} &    1.68\E{- 2} &    1.52\E{- 2} &    1.49\E{- 2} \\
\I{ 30}{Si} &    6.52\E{- 3} &    2.74\E{- 3} &    1.47\E{- 3} &    1.33\E{- 3} &    1.23\E{- 3} &    1.23\E{- 3} &    1.36\E{- 3} &    1.20\E{- 3} &    1.45\E{- 3} &    1.35\E{- 3} &    1.43\E{- 3} &    1.37\E{- 3} &    1.17\E{- 3} &    1.35\E{- 3} \\
\I{ 31}{ P} &    2.71\E{- 3} &    2.90\E{- 3} &    2.16\E{- 3} &    1.89\E{- 3} &    1.64\E{- 3} &    1.45\E{- 3} &    1.36\E{- 3} &    1.19\E{- 3} &    1.23\E{- 3} &    1.08\E{- 3} &    1.17\E{- 3} &    1.12\E{- 3} &    9.53\E{- 4} &    1.07\E{- 3} \\
\I{ 32}{ S} &    9.50\E{- 4} &    2.43\EE     &    4.12\EE     &    4.70\EE     &    6.08\EE     &    7.58\EE     &    8.85\EE     &    9.97\EE     &    1.08\E{  1} &    1.14\E{  1} &    1.18\E{  1} &    1.20\E{  1} &    1.18\E{  1} &    1.14\E{  1} \\
\I{ 33}{ S} &    3.84\E{- 5} &    3.07\E{- 3} &    3.50\E{- 3} &    3.64\E{- 3} &    3.73\E{- 3} &    3.75\E{- 3} &    3.74\E{- 3} &    3.68\E{- 3} &    3.59\E{- 3} &    3.47\E{- 3} &    3.34\E{- 3} &    3.18\E{- 3} &    2.96\E{- 3} &    2.71\E{- 3} \\
\I{ 34}{ S} &    2.06\E{- 4} &    6.36\E{- 3} &    4.57\E{- 3} &    2.94\E{- 3} &    1.93\E{- 3} &    1.22\E{- 3} &    7.83\E{- 4} &    4.98\E{- 4} &    3.28\E{- 4} &    2.12\E{- 4} &    1.49\E{- 4} &    1.08\E{- 4} &    8.83\E{- 5} &    9.78\E{- 5} \\
\I{ 36}{ S} &    1.12\E{- 8} &    1.12\E{- 8} &    7.86\E{- 9} &    6.60\E{- 9} &    5.53\E{- 9} &    4.55\E{- 9} &    3.82\E{- 9} &    3.24\E{- 9} &    2.73\E{- 9} &    2.36\E{- 9} &    2.05\E{- 9} &    1.79\E{- 9} &    1.59\E{- 9} &    1.49\E{- 9} \\
\I{ 35}{Cl} &    5.56\E{- 5} &    9.16\E{- 4} &    9.15\E{- 4} &    7.54\E{- 4} &    6.23\E{- 4} &    4.97\E{- 4} &    3.94\E{- 4} &    3.09\E{- 4} &    2.47\E{- 4} &    1.97\E{- 4} &    1.68\E{- 4} &    1.48\E{- 4} &    1.41\E{- 4} &    1.55\E{- 4} \\
\I{ 37}{Cl} &    1.05\E{- 7} &    5.59\E{- 4} &    6.41\E{- 4} &    6.69\E{- 4} &    6.85\E{- 4} &    6.90\E{- 4} &    6.86\E{- 4} &    6.76\E{- 4} &    6.55\E{- 4} &    6.32\E{- 4} &    6.08\E{- 4} &    5.78\E{- 4} &    5.36\E{- 4} &    4.91\E{- 4} \\
\I{ 36}{Ar} &    1.30\E{- 5} &    3.04\E{- 1} &    5.17\E{- 1} &    5.97\E{- 1} &    8.08\E{- 1} &    1.07\EE     &    1.32\EE     &    1.54\EE     &    1.72\EE     &    1.85\EE     &    1.94\EE     &    1.99\EE     &    1.99\EE     &    1.93\EE     \\
\I{ 38}{Ar} &    4.89\E{- 7} &    6.26\E{- 3} &    5.28\E{- 3} &    3.39\E{- 3} &    2.23\E{- 3} &    1.41\E{- 3} &    8.96\E{- 4} &    5.57\E{- 4} &    3.56\E{- 4} &    2.22\E{- 4} &    1.48\E{- 4} &    9.78\E{- 5} &    7.17\E{- 5} &    7.25\E{- 5} \\
\I{ 39}{ K} &    1.33\E{- 7} &    1.16\E{- 3} &    1.18\E{- 3} &    9.69\E{- 4} &    8.02\E{- 4} &    6.39\E{- 4} &    5.04\E{- 4} &    3.90\E{- 4} &    3.04\E{- 4} &    2.39\E{- 4} &    2.00\E{- 4} &    1.78\E{- 4} &    1.76\E{- 4} &    2.05\E{- 4} \\
\I{ 40}{ K} &    2.51\E{-10} &    3.49\E{- 8} &    3.02\E{- 8} &    2.19\E{- 8} &    1.69\E{- 8} &    1.28\E{- 8} &    9.67\E{- 9} &    7.29\E{- 9} &    5.46\E{- 9} &    4.05\E{- 9} &    3.17\E{- 9} &    2.54\E{- 9} &    2.27\E{- 9} &    2.60\E{- 9} \\
\I{ 41}{ K} &    2.45\E{-10} &    1.24\E{- 4} &    1.39\E{- 4} &    1.45\E{- 4} &    1.48\E{- 4} &    1.48\E{- 4} &    1.47\E{- 4} &    1.44\E{- 4} &    1.39\E{- 4} &    1.33\E{- 4} &    1.28\E{- 4} &    1.22\E{- 4} &    1.14\E{- 4} &    1.06\E{- 4} \\
\I{ 40}{Ca} &    1.46\E{- 8} &    1.88\E{- 1} &    3.17\E{- 1} &    3.70\E{- 1} &    5.29\E{- 1} &    7.61\E{- 1} &    9.93\E{- 1} &    1.22\EE     &    1.40\EE     &    1.54\EE     &    1.63\EE     &    1.69\EE     &    1.71\EE     &    1.67\EE     \\
\I{ 42}{Ca} &    2.39\E{-10} &    1.69\E{- 4} &    1.39\E{- 4} &    8.72\E{- 5} &    5.74\E{- 5} &    3.67\E{- 5} &    2.35\E{- 5} &    1.50\E{- 5} &    9.87\E{- 6} &    6.42\E{- 6} &    4.53\E{- 6} &    3.39\E{- 6} &    3.06\E{- 6} &    3.79\E{- 6} \\
\hline
\PPI&\PPE&\PPE&\PPE&\PPE&\PPE&\PPE&\PPE&\PPE&\PPE&\PPE&\PPE&\PPE&\PPE&\PPE \\
\end{tabular}
}\vspace{-1.5\baselineskip}
\begin{flushright}\textsc{(continued on next page)}\end{flushright}
\end{table}

\end{landscape}

\clearpage

\begin{landscape}

\addtocounter{table}{-1}

\begin{table}
\setlength{\tabcolsep}{1ex}
\centering
\caption{(continued) yields}
\scalebox{0.79}{
\begin{tabular}{r@{}lr@{}lr@{}lr@{}lr@{}lr@{}lr@{}lr@{}lr@{}lr@{}lr@{}lr@{}lr@{}lr@{}lr@{}l}
\hline
\hline
\multicolumn{2}{c}{ion} &
\multicolumn{2}{c}{ 65} &
\multicolumn{2}{c}{ 70} &
\multicolumn{2}{c}{ 75} &
\multicolumn{2}{c}{ 80} &
\multicolumn{2}{c}{ 85} &
\multicolumn{2}{c}{ 90} &
\multicolumn{2}{c}{ 95} &
\multicolumn{2}{c}{100} &
\multicolumn{2}{c}{105} &
\multicolumn{2}{c}{110} &
\multicolumn{2}{c}{115} &
\multicolumn{2}{c}{120} &
\multicolumn{2}{c}{125} &
\multicolumn{2}{c}{130} \\
\hline
\I{ 43}{Ca} &    3.56\E{-11} &    5.74\E{- 8} &    4.23\E{- 8} &    2.73\E{- 8} &    1.94\E{- 8} &    1.37\E{- 8} &    1.00\E{- 8} &    7.48\E{- 9} &    1.29\E{- 8} &    8.33\E{- 8} &    2.55\E{- 7} &    5.85\E{- 7} &    1.20\E{- 6} &    2.19\E{- 6} \\
\I{ 44}{Ca} &    6.96\E{-11} &    3.82\E{- 5} &    6.67\E{- 5} &    7.81\E{- 5} &    1.22\E{- 4} &    1.99\E{- 4} &    2.85\E{- 4} &    3.76\E{- 4} &    4.58\E{- 4} &    5.65\E{- 4} &    7.05\E{- 4} &    9.06\E{- 4} &    1.21\E{- 3} &    1.61\E{- 3} \\
\I{ 45}{Sc} &    1.40\E{-11} &    2.50\E{- 6} &    2.90\E{- 6} &    3.03\E{- 6} &    3.20\E{- 6} &    3.36\E{- 6} &    3.48\E{- 6} &    3.57\E{- 6} &    3.59\E{- 6} &    3.59\E{- 6} &    3.57\E{- 6} &    3.51\E{- 6} &    3.42\E{- 6} &    3.33\E{- 6} \\
\I{ 46}{Ti} &    2.29\E{-11} &    7.36\E{- 5} &    6.75\E{- 5} &    4.68\E{- 5} &    3.39\E{- 5} &    2.49\E{- 5} &    1.86\E{- 5} &    1.41\E{- 5} &    1.10\E{- 5} &    8.54\E{- 6} &    6.86\E{- 6} &    5.36\E{- 6} &    4.01\E{- 6} &    3.03\E{- 6} \\
\I{ 47}{Ti} &    5.08\E{-12} &    3.02\E{- 7} &    3.99\E{- 7} &    3.24\E{- 7} &    4.00\E{- 7} &    5.35\E{- 7} &    6.61\E{- 7} &    7.77\E{- 7} &    8.66\E{- 7} &    1.24\E{- 6} &    2.05\E{- 6} &    3.56\E{- 6} &    6.27\E{- 6} &    1.06\E{- 5} \\
\I{ 48}{Ti} &    6.41\E{-12} &    5.48\E{- 5} &    4.50\E{- 4} &    5.71\E{- 4} &    1.46\E{- 3} &    3.45\E{- 3} &    6.06\E{- 3} &    9.04\E{- 3} &    1.14\E{- 2} &    1.35\E{- 2} &    1.52\E{- 2} &    1.69\E{- 2} &    1.85\E{- 2} &    2.00\E{- 2} \\
\I{ 49}{Ti} &    5.16\E{-12} &    7.62\E{- 6} &    3.05\E{- 5} &    3.37\E{- 5} &    6.25\E{- 5} &    1.15\E{- 4} &    1.80\E{- 4} &    2.54\E{- 4} &    3.05\E{- 4} &    3.44\E{- 4} &    3.71\E{- 4} &    3.90\E{- 4} &    4.01\E{- 4} &    4.00\E{- 4} \\
\I{ 51}{ V} &    1.10\E{-12} &    1.74\E{- 5} &    6.49\E{- 5} &    6.54\E{- 5} &    1.11\E{- 4} &    1.92\E{- 4} &    2.95\E{- 4} &    4.17\E{- 4} &    4.76\E{- 4} &    5.04\E{- 4} &    5.15\E{- 4} &    5.16\E{- 4} &    5.09\E{- 4} &    4.95\E{- 4} \\
\I{ 50}{Cr} &    2.35\E{-13} &    3.15\E{- 4} &    4.65\E{- 4} &    3.84\E{- 4} &    3.99\E{- 4} &    4.34\E{- 4} &    4.87\E{- 4} &    5.65\E{- 4} &    5.68\E{- 4} &    5.42\E{- 4} &    5.15\E{- 4} &    4.85\E{- 4} &    4.51\E{- 4} &    4.20\E{- 4} \\
\I{ 52}{Cr} &    1.16\E{-12} &    7.01\E{- 4} &    6.16\E{- 3} &    7.76\E{- 3} &    2.24\E{- 2} &    6.24\E{- 2} &    1.25\E{- 1} &    2.06\E{- 1} &    2.67\E{- 1} &    3.16\E{- 1} &    3.52\E{- 1} &    3.79\E{- 1} &    3.97\E{- 1} &    4.00\E{- 1} \\
\I{ 53}{Cr} &    1.60\E{-13} &    1.23\E{- 4} &    5.75\E{- 4} &    6.36\E{- 4} &    1.32\E{- 3} &    2.84\E{- 3} &    5.14\E{- 3} &    8.34\E{- 3} &    1.04\E{- 2} &    1.19\E{- 2} &    1.29\E{- 2} &    1.36\E{- 2} &    1.40\E{- 2} &    1.40\E{- 2} \\
\I{ 54}{Cr} &    4.63\E{-13} &    1.90\E{- 8} &    1.75\E{- 8} &    1.13\E{- 8} &    7.36\E{- 9} &    5.23\E{- 9} &    4.31\E{- 9} &    4.19\E{- 9} &    3.54\E{- 9} &    2.84\E{- 9} &    2.33\E{- 9} &    1.91\E{- 9} &    1.58\E{- 9} &    1.31\E{- 9} \\
\I{ 55}{Mn} &    8.31\E{-14} &    7.02\E{- 4} &    2.75\E{- 3} &    2.91\E{- 3} &    5.43\E{- 3} &    1.11\E{- 2} &    2.02\E{- 2} &    3.49\E{- 2} &    4.26\E{- 2} &    4.63\E{- 2} &    4.82\E{- 2} &    4.85\E{- 2} &    4.85\E{- 2} &    4.75\E{- 2} \\
\I{ 54}{Fe} &    3.37\E{-15} &    2.46\E{- 2} &    4.49\E{- 2} &    4.06\E{- 2} &    4.94\E{- 2} &    6.44\E{- 2} &    8.79\E{- 2} &    1.30\E{- 1} &    1.42\E{- 1} &    1.41\E{- 1} &    1.38\E{- 1} &    1.31\E{- 1} &    1.24\E{- 1} &    1.18\E{- 1} \\
\I{ 56}{Fe} &    1.31\E{-13} &    1.17\E{- 2} &    1.02\E{- 1} &    1.29\E{- 1} &    4.07\E{- 1} &    1.31\EE     &    2.98\EE     &    5.82\EE     &    9.55\EE     &    1.42\E{  1} &    1.90\E{  1} &    2.46\E{  1} &    3.17\E{  1} &    3.96\E{  1} \\
\I{ 57}{Fe} &    2.80\E{-14} &    2.40\E{- 4} &    1.08\E{- 3} &    1.21\E{- 3} &    2.63\E{- 3} &    6.37\E{- 3} &    1.33\E{- 2} &    2.77\E{- 2} &    6.97\E{- 2} &    1.41\E{- 1} &    2.26\E{- 1} &    3.39\E{- 1} &    5.04\E{- 1} &    7.16\E{- 1} \\
\I{ 58}{Fe} &    5.78\E{-14} &    1.08\E{- 8} &    1.12\E{- 8} &    9.34\E{- 9} &    8.04\E{- 9} &    7.18\E{- 9} &    6.47\E{- 9} &    6.05\E{- 9} &    5.35\E{- 9} &    4.41\E{- 9} &    3.72\E{- 9} &    2.94\E{- 9} &    2.32\E{- 9} &    1.88\E{- 9} \\
\I{ 59}{Co} &    1.56\E{-14} &    3.67\E{- 6} &    4.09\E{- 6} &    4.34\E{- 6} &    4.60\E{- 6} &    5.05\E{- 6} &    5.59\E{- 6} &    1.31\E{- 5} &    6.67\E{- 4} &    2.28\E{- 3} &    4.47\E{- 3} &    7.62\E{- 3} &    1.25\E{- 2} &    1.93\E{- 2} \\
\I{ 58}{Ni} &    6.40\E{-17} &    1.34\E{- 3} &    2.48\E{- 3} &    2.46\E{- 3} &    3.36\E{- 3} &    5.11\E{- 3} &    8.24\E{- 3} &    1.84\E{- 2} &    8.66\E{- 2} &    2.07\E{- 1} &    3.59\E{- 1} &    5.77\E{- 1} &    9.30\E{- 1} &    1.47\EE     \\
\I{ 60}{Ni} &    1.58\E{-14} &    7.16\E{- 6} &    7.18\E{- 6} &    6.73\E{- 6} &    6.24\E{- 6} &    5.94\E{- 6} &    5.55\E{- 6} &    3.69\E{- 5} &    9.03\E{- 3} &    3.67\E{- 2} &    7.61\E{- 2} &    1.32\E{- 1} &    2.13\E{- 1} &    3.13\E{- 1} \\
\I{ 61}{Ni} &    4.68\E{-15} &    8.05\E{-10} &    8.87\E{-10} &    9.11\E{-10} &    9.40\E{-10} &    9.90\E{-10} &    1.04\E{- 9} &    1.91\E{- 6} &    5.42\E{- 4} &    2.22\E{- 3} &    4.60\E{- 3} &    7.99\E{- 3} &    1.29\E{- 2} &    1.92\E{- 2} \\
\I{ 62}{Ni} &    2.05\E{-14} &    1.47\E{-10} &    2.18\E{-10} &    1.97\E{-10} &    2.45\E{-10} &    3.62\E{-10} &    6.33\E{-10} &    7.97\E{- 6} &    2.65\E{- 3} &    1.15\E{- 2} &    2.51\E{- 2} &    4.59\E{- 2} &    8.04\E{- 2} &    1.33\E{- 1} \\
\I{ 63}{Cu} &    2.55\E{-15} &    3.92\E{-12} &    4.33\E{-12} &    4.78\E{-12} &    5.04\E{-12} &    5.55\E{-12} &    6.41\E{-12} &    2.90\E{- 9} &    1.49\E{- 6} &    7.14\E{- 6} &    1.62\E{- 5} &    3.04\E{- 5} &    5.39\E{- 5} &    8.93\E{- 5} \\
\I{ 65}{Cu} &    3.35\E{-15} &    8.19\E{-15} &    1.01\E{-14} &    1.75\E{-14} &    1.89\E{-14} &    1.80\E{-14} &    1.48\E{-14} &    3.06\E{-10} &    7.71\E{- 7} &    4.47\E{- 6} &    1.07\E{- 5} &    2.04\E{- 5} &    3.53\E{- 5} &    5.50\E{- 5} \\
\I{ 64}{Zn} &    1.39\E{-15} &    1.68\E{-11} &    1.93\E{-11} &    2.32\E{-11} &    2.58\E{-11} &    2.74\E{-11} &    2.86\E{-11} &    4.22\E{- 9} &    8.97\E{- 6} &    5.11\E{- 5} &    1.21\E{- 4} &    2.29\E{- 4} &    3.93\E{- 4} &    6.00\E{- 4} \\
\I{ 66}{Zn} &    3.22\E{-15} &    8.57\E{-15} &    1.08\E{-14} &    2.28\E{-14} &    2.23\E{-14} &    1.71\E{-14} &    8.52\E{-15} &    5.50\E{- 9} &    1.40\E{- 5} &    8.43\E{- 5} &    2.07\E{- 4} &    4.09\E{- 4} &    7.52\E{- 4} &    1.28\E{- 3} \\
\I{ 67}{Zn} &    6.27\E{-16} &    4.18\E{-16} &    4.13\E{-16} &    4.69\E{-16} &    3.13\E{-16} &    4.92\E{-16} &    6.63\E{-17} &    3.86\E{-12} &    1.13\E{- 8} &    7.11\E{- 8} &    1.80\E{- 7} &    3.63\E{- 7} &    6.85\E{- 7} &    1.20\E{- 6} \\
\I{ 68}{Zn} &    3.58\E{-15} &    1.85\E{-15} &    1.80\E{-15} &    2.02\E{-15} &    1.26\E{-15} &    2.18\E{-15} &    2.72\E{-16} &    9.36\E{-13} &    3.31\E{- 9} &    2.30\E{- 8} &    6.12\E{- 8} &    1.29\E{- 7} &    2.58\E{- 7} &    4.82\E{- 7} \\
\I{ 69}{Ga} &    4.96\E{-16} &    4.19\E{-16} &    3.86\E{-16} &    4.20\E{-16} &    2.29\E{-16} &    4.07\E{-16} &    2.60\E{-17} &    2.72\E{-14} &    3.41\E{-10} &    2.63\E{- 9} &    7.11\E{- 9} &    1.47\E{- 8} &    2.75\E{- 8} &    4.58\E{- 8} \\
\I{ 71}{Ga} &    4.53\E{-16} &    2.19\E{-16} &    2.00\E{-16} &    2.19\E{-16} &    1.15\E{-16} &    2.06\E{-16} &    1.15\E{-17} &    8.19\E{-15} &    2.34\E{-10} &    2.00\E{- 9} &    5.59\E{- 9} &    1.18\E{- 8} &    2.27\E{- 8} &    3.94\E{- 8} \\
\I{ 70}{Ge} &    6.63\E{-16} &    1.33\E{-15} &    1.43\E{-15} &    1.61\E{-15} &    9.13\E{-16} &    1.70\E{-15} &    1.08\E{-16} &    4.04\E{-12} &    9.27\E{- 8} &    7.68\E{- 7} &    2.12\E{- 6} &    4.43\E{- 6} &    8.46\E{- 6} &    1.46\E{- 5} \\
\I{ 72}{Ge} &    7.04\E{-16} &    4.76\E{-16} &    4.11\E{-16} &    4.36\E{-16} &    2.17\E{-16} &    3.87\E{-16} &    2.04\E{-17} &    3.35\E{-15} &    1.85\E{-10} &    1.79\E{- 9} &    5.37\E{- 9} &    1.20\E{- 8} &    2.50\E{- 8} &    4.81\E{- 8} \\
\I{ 74}{Se} &    2.74\E{-19} &    2.34\E{-16} &    3.16\E{-16} &    3.83\E{-16} &    2.24\E{-16} &    4.61\E{-16} &    2.49\E{-17} &    6.27\E{-16} &    2.18\E{-10} &    2.54\E{- 9} &    7.86\E{- 9} &    1.74\E{- 8} &    3.37\E{- 8} &    5.75\E{- 8} \\
\hline
\PPI&\PPE&\PPE&\PPE&\PPE&\PPE&\PPE&\PPE&\PPE&\PPE&\PPE&\PPE&\PPE&\PPE&\PPE \\
\end{tabular}
}\vspace{-1.5\baselineskip}
\begin{flushright}\textsc{(end of yield table)}\end{flushright}
\end{table}

\end{landscape}

\clearpage

\begin{landscape}


\begin{table}
\setlength{\tabcolsep}{1ex}
\centering
\caption{production factors (relative to solar)\lTab{pf}}
\scalebox{0.79}{
\begin{tabular}{r@{}lr@{}lr@{}lr@{}lr@{}lr@{}lr@{}lr@{}lr@{}lr@{}lr@{}lr@{}lr@{}lr@{}lr@{}l}
\hline
\hline
\multicolumn{2}{c}{ion} &
\multicolumn{2}{c}{ 65} &
\multicolumn{2}{c}{ 70} &
\multicolumn{2}{c}{ 75} &
\multicolumn{2}{c}{ 80} &
\multicolumn{2}{c}{ 85} &
\multicolumn{2}{c}{ 90} &
\multicolumn{2}{c}{ 95} &
\multicolumn{2}{c}{100} &
\multicolumn{2}{c}{105} &
\multicolumn{2}{c}{110} &
\multicolumn{2}{c}{115} &
\multicolumn{2}{c}{120} &
\multicolumn{2}{c}{125} &
\multicolumn{2}{c}{130} \\
\hline
\I{  4}{He} &    1.10\E{- 1} &    8.05\E{- 2} &    6.97\E{- 2} &    6.61\E{- 2} &    6.15\E{- 2} &    5.70\E{- 2} &    5.45\E{- 2} &    5.33\E{- 2} &    5.28\E{- 2} &    5.48\E{- 2} &    5.61\E{- 2} &    6.04\E{- 2} &    6.91\E{- 2} &    7.89\E{- 2} \\
\I{ 11}{ B} &    2.38\E{- 7} &    2.35\E{- 8} &    8.04\E{- 7} &    1.79\E{- 6} &    3.55\E{- 6} &    5.29\E{- 6} &    6.87\E{- 6} &    8.42\E{- 6} &    9.73\E{- 6} &    1.08\E{- 5} &    1.18\E{- 5} &    1.23\E{- 5} &    1.26\E{- 5} &    1.24\E{- 5} \\
\I{ 12}{ C} &    3.50\E{  1} &    2.14\E{  1} &    1.90\E{  1} &    1.79\E{  1} &    1.66\E{  1} &    1.54\E{  1} &    1.43\E{  1} &    1.32\E{  1} &    1.21\E{  1} &    1.12\E{  1} &    1.07\E{  1} &    1.02\E{  1} &    9.53\EE     &    8.85\EE     \\
\I{ 13}{ C} &    9.84\E{- 5} &    1.64\E{- 4} &    2.72\E{- 4} &    3.04\E{- 4} &    2.74\E{- 4} &    1.76\E{- 4} &    1.03\E{- 4} &    1.18\E{- 4} &    6.00\E{- 5} &    3.98\E{- 5} &    3.91\E{- 5} &    3.45\E{- 5} &    3.03\E{- 5} &    1.97\E{- 5} \\
\I{ 14}{ N} &    1.09\E{- 3} &    6.51\E{- 4} &    5.20\E{- 4} &    4.70\E{- 4} &    4.03\E{- 4} &    3.48\E{- 4} &    3.07\E{- 4} &    2.80\E{- 4} &    2.52\E{- 4} &    2.26\E{- 4} &    1.91\E{- 4} &    1.47\E{- 4} &    1.02\E{- 4} &    7.65\E{- 5} \\
\I{ 15}{ N} &    2.56\E{- 3} &    2.30\E{- 2} &    2.05\E{- 2} &    1.92\E{- 2} &    1.77\E{- 2} &    1.64\E{- 2} &    1.55\E{- 2} &    1.47\E{- 2} &    1.42\E{- 2} &    1.41\E{- 2} &    1.39\E{- 2} &    1.37\E{- 2} &    1.29\E{- 2} &    1.17\E{- 2} \\
\I{ 16}{ O} &    7.90\E{  1} &    6.83\E{  1} &    6.18\E{  1} &    6.10\E{  1} &    5.72\E{  1} &    5.32\E{  1} &    4.96\E{  1} &    4.58\E{  1} &    4.24\E{  1} &    3.90\E{  1} &    3.60\E{  1} &    3.31\E{  1} &    2.99\E{  1} &    2.68\E{  1} \\
\I{ 17}{ O} &    1.69\E{- 2} &    3.62\E{- 3} &    2.66\E{- 3} &    2.40\E{- 3} &    2.00\E{- 3} &    1.70\E{- 3} &    1.46\E{- 3} &    1.31\E{- 3} &    1.16\E{- 3} &    1.04\E{- 3} &    9.24\E{- 4} &    8.02\E{- 4} &    6.70\E{- 4} &    5.66\E{- 4} \\
\I{ 18}{ O} &    1.99\E{- 3} &    6.55\E{- 4} &    5.23\E{- 4} &    4.70\E{- 4} &    4.32\E{- 4} &    4.02\E{- 4} &    3.71\E{- 4} &    3.74\E{- 4} &    3.51\E{- 4} &    3.45\E{- 4} &    3.35\E{- 4} &    3.27\E{- 4} &    3.28\E{- 4} &    3.01\E{- 4} \\
\I{ 19}{ F} &    6.20\E{- 4} &    4.96\E{- 4} &    4.23\E{- 4} &    3.79\E{- 4} &    3.59\E{- 4} &    3.90\E{- 4} &    4.18\E{- 4} &    4.17\E{- 4} &    4.54\E{- 4} &    4.56\E{- 4} &    5.15\E{- 4} &    5.40\E{- 4} &    5.46\E{- 4} &    4.89\E{- 4} \\
\I{ 20}{Ne} &    4.75\E{  1} &    3.56\E{  1} &    3.20\E{  1} &    3.26\E{  1} &    3.06\E{  1} &    2.82\E{  1} &    2.59\E{  1} &    2.51\E{  1} &    2.27\E{  1} &    2.19\E{  1} &    2.06\E{  1} &    2.00\E{  1} &    1.94\E{  1} &    1.77\E{  1} \\
\I{ 21}{Ne} &    1.88\EE     &    2.26\E{- 1} &    6.78\E{- 1} &    7.55\E{- 1} &    8.53\E{- 1} &    6.77\E{- 1} &    4.64\E{- 1} &    5.18\E{- 1} &    3.28\E{- 1} &    3.64\E{- 1} &    3.06\E{- 1} &    3.07\E{- 1} &    3.26\E{- 1} &    2.32\E{- 1} \\
\I{ 22}{Ne} &    2.20\E{- 2} &    2.20\E{- 3} &    1.90\E{- 3} &    1.89\E{- 3} &    1.98\E{- 3} &    1.94\E{- 3} &    1.83\E{- 3} &    1.82\E{- 3} &    1.67\E{- 3} &    1.72\E{- 3} &    1.60\E{- 3} &    1.54\E{- 3} &    1.43\E{- 3} &    1.19\E{- 3} \\
\I{ 23}{Na} &    4.21\EE     &    1.34\EE     &    1.18\EE     &    1.13\EE     &    1.04\EE     &    9.58\E{- 1} &    8.88\E{- 1} &    8.27\E{- 1} &    7.69\E{- 1} &    7.26\E{- 1} &    6.97\E{- 1} &    6.85\E{- 1} &    6.72\E{- 1} &    6.41\E{- 1} \\
\I{ 24}{Mg} &    4.57\E{  1} &    8.38\E{  1} &    9.03\E{  1} &    8.92\E{  1} &    9.08\E{  1} &    9.16\E{  1} &    8.95\E{  1} &    8.57\E{  1} &    8.15\E{  1} &    7.61\E{  1} &    7.60\E{  1} &    7.37\E{  1} &    6.87\E{  1} &    6.54\E{  1} \\
\I{ 25}{Mg} &    8.61\E{- 1} &    2.28\E{- 1} &    6.06\E{- 1} &    4.12\E{- 1} &    5.85\E{- 1} &    6.81\E{- 1} &    7.32\E{- 1} &    6.32\E{- 1} &    6.43\E{- 1} &    6.11\E{- 1} &    5.19\E{- 1} &    4.63\E{- 1} &    4.01\E{- 1} &    4.05\E{- 1} \\
\I{ 26}{Mg} &    7.36\E{- 1} &    3.20\E{- 1} &    2.90\E{- 1} &    2.80\E{- 1} &    2.72\E{- 1} &    2.75\E{- 1} &    2.80\E{- 1} &    2.47\E{- 1} &    2.52\E{- 1} &    2.23\E{- 1} &    2.17\E{- 1} &    2.00\E{- 1} &    1.77\E{- 1} &    1.71\E{- 1} \\
\I{ 27}{Al} &    8.95\EE     &    4.36\EE     &    3.82\EE     &    3.43\EE     &    3.39\EE     &    3.30\EE     &    3.21\EE     &    2.81\EE     &    2.73\EE     &    2.42\EE     &    2.25\EE     &    2.03\EE     &    1.78\EE     &    1.78\EE     \\
\I{ 28}{Si} &    7.43\EE     &    1.74\E{  2} &    2.50\E{  2} &    2.60\E{  2} &    2.97\E{  2} &    3.27\E{  2} &    3.45\E{  2} &    3.53\E{  2} &    3.57\E{  2} &    3.51\E{  2} &    3.42\E{  2} &    3.28\E{  2} &    3.08\E{  2} &    2.85\E{  2} \\
\I{ 29}{Si} &    8.80\E{- 1} &    9.69\EE     &    8.74\EE     &    8.11\EE     &    7.35\EE     &    6.72\EE     &    6.30\EE     &    5.67\EE     &    5.39\EE     &    4.80\EE     &    4.52\EE     &    4.09\EE     &    3.55\EE     &    3.35\EE     \\
\I{ 30}{Si} &    4.27\EE     &    1.66\EE     &    8.36\E{- 1} &    7.07\E{- 1} &    6.16\E{- 1} &    5.81\E{- 1} &    6.07\E{- 1} &    5.08\E{- 1} &    5.88\E{- 1} &    5.24\E{- 1} &    5.30\E{- 1} &    4.87\E{- 1} &    3.99\E{- 1} &    4.40\E{- 1} \\
\I{ 31}{ P} &    5.12\EE     &    5.07\EE     &    3.54\EE     &    2.90\EE     &    2.37\EE     &    1.98\EE     &    1.75\EE     &    1.46\EE     &    1.44\EE     &    1.21\EE     &    1.25\EE     &    1.14\EE     &    9.36\E{- 1} &    1.01\EE     \\
\I{ 32}{ S} &    3.69\E{- 2} &    8.77\E{  1} &    1.39\E{  2} &    1.48\E{  2} &    1.81\E{  2} &    2.13\E{  2} &    2.36\E{  2} &    2.52\E{  2} &    2.61\E{  2} &    2.63\E{  2} &    2.59\E{  2} &    2.52\E{  2} &    2.40\E{  2} &    2.22\E{  2} \\
\I{ 33}{ S} &    1.84\E{- 1} &    1.36\E{  1} &    1.45\E{  1} &    1.41\E{  1} &    1.36\E{  1} &    1.29\E{  1} &    1.22\E{  1} &    1.14\E{  1} &    1.06\E{  1} &    9.78\EE     &    9.03\EE     &    8.24\EE     &    7.36\EE     &    6.47\EE     \\
\I{ 34}{ S} &    1.70\E{- 1} &    4.87\EE     &    3.27\EE     &    1.97\EE     &    1.21\EE     &    7.27\E{- 1} &    4.42\E{- 1} &    2.67\E{- 1} &    1.68\E{- 1} &    1.03\E{- 1} &    6.94\E{- 2} &    4.82\E{- 2} &    3.79\E{- 2} &    4.03\E{- 2} \\
\I{ 36}{ S} &    1.84\E{- 3} &    1.70\E{- 3} &    1.12\E{- 3} &    8.80\E{- 4} &    6.95\E{- 4} &    5.39\E{- 4} &    4.29\E{- 4} &    3.45\E{- 4} &    2.78\E{- 4} &    2.29\E{- 4} &    1.90\E{- 4} &    1.60\E{- 4} &    1.35\E{- 4} &    1.22\E{- 4} \\
\I{ 35}{Cl} &    3.38\E{- 1} &    5.17\EE     &    4.82\EE     &    3.73\EE     &    2.90\EE     &    2.18\EE     &    1.64\EE     &    1.22\EE     &    9.29\E{- 1} &    7.09\E{- 1} &    5.77\E{- 1} &    4.87\E{- 1} &    4.45\E{- 1} &    4.72\E{- 1} \\
\I{ 37}{Cl} &    1.88\E{- 3} &    9.35\EE     &    1.00\E{  1} &    9.79\EE     &    9.44\EE     &    8.98\EE     &    8.45\EE     &    7.91\EE     &    7.31\EE     &    6.73\EE     &    6.19\EE     &    5.64\EE     &    5.03\EE     &    4.42\EE     \\
\I{ 36}{Ar} &    2.59\E{- 3} &    5.62\E{  1} &    8.92\E{  1} &    9.64\E{  1} &    1.23\E{  2} &    1.54\E{  2} &    1.79\E{  2} &    1.99\E{  2} &    2.11\E{  2} &    2.17\E{  2} &    2.18\E{  2} &    2.14\E{  2} &    2.05\E{  2} &    1.92\E{  2} \\
\I{ 38}{Ar} &    4.90\E{- 4} &    5.82\EE     &    4.58\EE     &    2.76\EE     &    1.71\EE     &    1.02\EE     &    6.13\E{- 1} &    3.62\E{- 1} &    2.21\E{- 1} &    1.31\E{- 1} &    8.34\E{- 2} &    5.30\E{- 2} &    3.73\E{- 2} &    3.63\E{- 2} \\
\I{ 40}{Ar} &    4.64\E{- 5} &    3.84\E{- 5} &    2.30\E{- 5} &    1.71\E{- 5} &    1.57\E{- 5} &    1.21\E{- 5} &    1.20\E{- 5} &    1.18\E{- 5} &    8.77\E{- 6} &    8.30\E{- 6} &    1.12\E{- 5} &    1.06\E{- 5} &    1.31\E{- 5} &    1.74\E{- 5} \\
\I{ 39}{ K} &    5.89\E{- 4} &    4.77\EE     &    4.52\EE     &    3.49\EE     &    2.72\EE     &    2.05\EE     &    1.53\EE     &    1.12\EE     &    8.36\E{- 1} &    6.26\E{- 1} &    5.02\E{- 1} &    4.26\E{- 1} &    4.05\E{- 1} &    4.55\E{- 1} \\
\I{ 40}{ K} &    6.96\E{- 4} &    8.99\E{- 2} &    7.27\E{- 2} &    4.93\E{- 2} &    3.60\E{- 2} &    2.56\E{- 2} &    1.84\E{- 2} &    1.32\E{- 2} &    9.38\E{- 3} &    6.65\E{- 3} &    4.97\E{- 3} &    3.81\E{- 3} &    3.28\E{- 3} &    3.61\E{- 3} \\
\I{ 41}{ K} &    1.43\E{- 5} &    6.72\EE     &    7.05\EE     &    6.88\EE     &    6.60\EE     &    6.26\EE     &    5.87\EE     &    5.47\EE     &    5.02\EE     &    4.61\EE     &    4.23\EE     &    3.85\EE     &    3.46\EE     &    3.09\EE     \\
\I{ 40}{Ca} &    3.75\E{- 6} &    4.48\E{  1} &    7.06\E{  1} &    7.72\E{  1} &    1.04\E{  2} &    1.41\E{  2} &    1.74\E{  2} &    2.03\E{  2} &    2.22\E{  2} &    2.33\E{  2} &    2.37\E{  2} &    2.36\E{  2} &    2.29\E{  2} &    2.15\E{  2} \\
\I{ 42}{Ca} &    8.77\E{- 6} &    5.75\EE     &    4.42\EE     &    2.60\EE     &    1.61\EE     &    9.72\E{- 1} &    5.91\E{- 1} &    3.59\E{- 1} &    2.24\E{- 1} &    1.39\E{- 1} &    9.40\E{- 2} &    6.74\E{- 2} &    5.85\E{- 2} &    6.94\E{- 2} \\
\I{ 43}{Ca} &    6.11\E{- 6} &    9.15\E{- 3} &    6.29\E{- 3} &    3.80\E{- 3} &    2.54\E{- 3} &    1.70\E{- 3} &    1.18\E{- 3} &    8.34\E{- 4} &    1.37\E{- 3} &    8.45\E{- 3} &    2.47\E{- 2} &    5.43\E{- 2} &    1.07\E{- 1} &    1.88\E{- 1} \\
\I{ 44}{Ca} &    7.52\E{- 7} &    3.83\E{- 1} &    6.24\E{- 1} &    6.85\E{- 1} &    1.00\EE     &    1.55\EE     &    2.10\EE     &    2.64\EE     &    3.06\EE     &    3.61\EE     &    4.30\EE     &    5.30\EE     &    6.79\EE     &    8.67\EE     \\
\I{ 46}{Ca} &    3.11\E{- 7} &    5.30\E{- 8} &    4.66\E{- 9} &    2.95\E{- 9} &    2.01\E{- 9} &    1.40\E{- 9} &    7.63\E{-10} &    5.98\E{-10} &    8.11\E{-10} &    7.43\E{-10} &    3.79\E{-10} &    3.99\E{-10} &    3.50\E{-10} &    3.35\E{-10} \\
\I{ 45}{Sc} &    5.55\E{- 6} &    9.19\E{- 1} &    9.93\E{- 1} &    9.74\E{- 1} &    9.67\E{- 1} &    9.60\E{- 1} &    9.40\E{- 1} &    9.18\E{- 1} &    8.79\E{- 1} &    8.38\E{- 1} &    7.97\E{- 1} &    7.53\E{- 1} &    7.04\E{- 1} &    6.59\E{- 1} \\
\I{ 46}{Ti} &    1.58\E{- 6} &    4.71\EE     &    4.03\EE     &    2.62\EE     &    1.79\EE     &    1.24\EE     &    8.75\E{- 1} &    6.33\E{- 1} &    4.69\E{- 1} &    3.48\E{- 1} &    2.67\E{- 1} &    2.00\E{- 1} &    1.44\E{- 1} &    1.04\E{- 1} \\
\I{ 47}{Ti} &    3.76\E{- 7} &    2.08\E{- 2} &    2.56\E{- 2} &    1.94\E{- 2} &    2.26\E{- 2} &    2.85\E{- 2} &    3.35\E{- 2} &    3.74\E{- 2} &    3.96\E{- 2} &    5.42\E{- 2} &    8.58\E{- 2} &    1.42\E{- 1} &    2.41\E{- 1} &    3.90\E{- 1} \\
\I{ 48}{Ti} &    4.59\E{- 8} &    3.65\E{- 1} &    2.80\EE     &    3.32\EE     &    7.97\EE     &    1.78\E{  1} &    2.97\E{  1} &    4.21\E{  1} &    5.06\E{  1} &    5.71\E{  1} &    6.17\E{  1} &    6.55\E{  1} &    6.90\E{  1} &    7.15\E{  1} \\
\hline
\PPI&\PPE&\PPE&\PPE&\PPE&\PPE&\PPE&\PPE&\PPE&\PPE&\PPE&\PPE&\PPE&\PPE&\PPE \\
\end{tabular}
}\vspace{-1.5\baselineskip}
\begin{flushright}\textsc{(continued on next page)}\end{flushright}
\end{table}

\end{landscape}

\clearpage

\begin{landscape}

\addtocounter{table}{-1}

\begin{table}
\setlength{\tabcolsep}{1ex}
\centering
\caption{(continued) production factors}
\scalebox{0.79}{
\begin{tabular}{r@{}lr@{}lr@{}lr@{}lr@{}lr@{}lr@{}lr@{}lr@{}lr@{}lr@{}lr@{}lr@{}lr@{}lr@{}l}
\hline
\hline
\multicolumn{2}{c}{ion} &
\multicolumn{2}{c}{ 65} &
\multicolumn{2}{c}{ 70} &
\multicolumn{2}{c}{ 75} &
\multicolumn{2}{c}{ 80} &
\multicolumn{2}{c}{ 85} &
\multicolumn{2}{c}{ 90} &
\multicolumn{2}{c}{ 95} &
\multicolumn{2}{c}{100} &
\multicolumn{2}{c}{105} &
\multicolumn{2}{c}{110} &
\multicolumn{2}{c}{115} &
\multicolumn{2}{c}{120} &
\multicolumn{2}{c}{125} &
\multicolumn{2}{c}{130} \\
\hline
\I{ 49}{Ti} &    4.85\E{- 7} &    6.65\E{- 1} &    2.49\EE     &    2.58\EE     &    4.49\EE     &    7.82\EE     &    1.16\E{  1} &    1.55\E{  1} &    1.78\E{  1} &    1.91\E{  1} &    1.97\E{  1} &    1.99\E{  1} &    1.96\E{  1} &    1.88\E{  1} \\
\I{ 50}{Ti} &    6.43\E{- 7} &    3.53\E{- 7} &    2.02\E{- 7} &    1.82\E{- 7} &    1.33\E{- 7} &    1.39\E{- 7} &    5.71\E{- 8} &    3.95\E{- 8} &    1.01\E{- 7} &    9.07\E{- 8} &    2.29\E{- 8} &    4.34\E{- 8} &    3.61\E{- 8} &    2.28\E{- 8} \\
\I{ 50}{ V} &    6.42\E{- 6} &    3.08\E{- 3} &    1.95\E{- 3} &    9.64\E{- 4} &    5.40\E{- 4} &    3.15\E{- 4} &    1.98\E{- 4} &    1.34\E{- 4} &    9.30\E{- 5} &    6.29\E{- 5} &    4.38\E{- 5} &    3.12\E{- 5} &    2.20\E{- 5} &    1.65\E{- 5} \\
\I{ 51}{ V} &    4.48\E{- 8} &    6.58\E{- 1} &    2.30\EE     &    2.17\EE     &    3.45\EE     &    5.67\EE     &    8.24\EE     &    1.11\E{  1} &    1.20\E{  1} &    1.22\E{  1} &    1.19\E{  1} &    1.14\E{  1} &    1.08\E{  1} &    1.01\E{  1} \\
\I{ 50}{Cr} &    4.87\E{- 9} &    6.06\EE     &    8.35\EE     &    6.47\EE     &    6.32\EE     &    6.50\EE     &    6.91\EE     &    7.61\EE     &    7.29\EE     &    6.64\EE     &    6.04\EE     &    5.44\EE     &    4.86\EE     &    4.36\EE     \\
\I{ 52}{Cr} &    1.20\E{- 9} &    6.74\E{- 1} &    5.53\EE     &    6.53\EE     &    1.77\E{  1} &    4.67\E{  1} &    8.83\E{  1} &    1.39\E{  2} &    1.71\E{  2} &    1.94\E{  2} &    2.06\E{  2} &    2.13\E{  2} &    2.14\E{  2} &    2.07\E{  2} \\
\I{ 53}{Cr} &    1.44\E{- 9} &    1.02\EE     &    4.47\EE     &    4.63\EE     &    9.03\EE     &    1.84\E{  1} &    3.15\E{  1} &    4.86\E{  1} &    5.79\E{  1} &    6.30\E{  1} &    6.54\E{  1} &    6.61\E{  1} &    6.54\E{  1} &    6.29\E{  1} \\
\I{ 54}{Cr} &    1.64\E{- 8} &    6.22\E{- 4} &    5.34\E{- 4} &    3.25\E{- 4} &    1.99\E{- 4} &    1.33\E{- 4} &    1.04\E{- 4} &    9.62\E{- 5} &    7.75\E{- 5} &    5.92\E{- 5} &    4.66\E{- 5} &    3.66\E{- 5} &    2.90\E{- 5} &    2.32\E{- 5} \\
\I{ 55}{Mn} &    9.62\E{-11} &    7.55\E{- 1} &    2.76\EE     &    2.73\EE     &    4.81\EE     &    9.25\EE     &    1.60\E{  1} &    2.63\E{  1} &    3.05\E{  1} &    3.17\E{  1} &    3.15\E{  1} &    3.04\E{  1} &    2.92\E{  1} &    2.75\E{  1} \\
\I{ 54}{Fe} &    7.28\E{-13} &    4.94\EE     &    8.40\EE     &    7.12\EE     &    8.16\EE     &    1.00\E{  1} &    1.30\E{  1} &    1.82\E{  1} &    1.90\E{  1} &    1.80\E{  1} &    1.68\E{  1} &    1.53\E{  1} &    1.39\E{  1} &    1.27\E{  1} \\
\I{ 56}{Fe} &    1.73\E{-12} &    1.43\E{- 1} &    1.16\EE     &    1.38\EE     &    4.10\EE     &    1.24\E{  1} &    2.69\E{  1} &    4.98\E{  1} &    7.79\E{  1} &    1.11\E{  2} &    1.41\E{  2} &    1.75\E{  2} &    2.17\E{  2} &    2.61\E{  2} \\
\I{ 57}{Fe} &    1.51\E{-11} &    1.20\E{- 1} &    5.03\E{- 1} &    5.28\E{- 1} &    1.08\EE     &    2.48\EE     &    4.91\EE     &    9.69\EE     &    2.32\E{  1} &    4.48\E{  1} &    6.88\E{  1} &    9.91\E{  1} &    1.41\E{  2} &    1.93\E{  2} \\
\I{ 58}{Fe} &    2.41\E{-10} &    4.18\E{- 5} &    4.03\E{- 5} &    3.16\E{- 5} &    2.56\E{- 5} &    2.16\E{- 5} &    1.84\E{- 5} &    1.64\E{- 5} &    1.38\E{- 5} &    1.08\E{- 5} &    8.75\E{- 6} &    6.62\E{- 6} &    5.02\E{- 6} &    3.92\E{- 6} \\
\I{ 59}{Co} &    7.14\E{-11} &    1.56\E{- 2} &    1.62\E{- 2} &    1.62\E{- 2} &    1.61\E{- 2} &    1.67\E{- 2} &    1.75\E{- 2} &    3.91\E{- 2} &    1.89\EE     &    6.18\EE     &    1.16\E{  1} &    1.89\E{  1} &    2.98\E{  1} &    4.43\E{  1} \\
\I{ 58}{Ni} &    1.99\E{-14} &    3.88\E{- 1} &    6.70\E{- 1} &    6.22\E{- 1} &    8.00\E{- 1} &    1.15\EE     &    1.75\EE     &    3.71\EE     &    1.67\E{  1} &    3.81\E{  1} &    6.31\E{  1} &    9.72\E{  1} &    1.50\E{  2} &    2.29\E{  2} \\
\I{ 60}{Ni} &    1.24\E{-11} &    5.23\E{- 3} &    4.89\E{- 3} &    4.30\E{- 3} &    3.75\E{- 3} &    3.37\E{- 3} &    2.98\E{- 3} &    1.89\E{- 2} &    4.39\EE     &    1.71\E{  1} &    3.38\E{  1} &    5.61\E{  1} &    8.70\E{  1} &    1.23\E{  2} \\
\I{ 61}{Ni} &    8.39\E{-11} &    1.34\E{- 5} &    1.38\E{- 5} &    1.33\E{- 5} &    1.29\E{- 5} &    1.28\E{- 5} &    1.28\E{- 5} &    2.22\E{- 2} &    6.01\EE     &    2.35\E{  1} &    4.66\E{  1} &    7.75\E{  1} &    1.21\E{  2} &    1.72\E{  2} \\
\I{ 62}{Ni} &    1.13\E{-10} &    7.56\E{- 7} &    1.05\E{- 6} &    8.89\E{- 7} &    1.04\E{- 6} &    1.45\E{- 6} &    2.40\E{- 6} &    2.87\E{- 2} &    9.11\EE     &    3.78\E{  1} &    7.85\E{  1} &    1.38\E{  2} &    2.32\E{  2} &    3.68\E{  2} \\
\I{ 63}{Cu} &    6.81\E{-11} &    9.74\E{- 8} &    1.00\E{- 7} &    1.04\E{- 7} &    1.03\E{- 7} &    1.07\E{- 7} &    1.17\E{- 7} &    5.04\E{- 5} &    2.47\E{- 2} &    1.13\E{- 1} &    2.44\E{- 1} &    4.40\E{- 1} &    7.51\E{- 1} &    1.20\EE     \\
\I{ 65}{Cu} &    1.95\E{-10} &    4.42\E{-10} &    5.09\E{-10} &    8.24\E{-10} &    8.41\E{-10} &    7.55\E{-10} &    5.89\E{-10} &    1.16\E{- 5} &    2.78\E{- 2} &    1.54\E{- 1} &    3.51\E{- 1} &    6.42\E{- 1} &    1.07\EE     &    1.60\EE     \\
\I{ 64}{Zn} &    2.16\E{-11} &    2.43\E{- 7} &    2.59\E{- 7} &    2.92\E{- 7} &    3.06\E{- 7} &    3.07\E{- 7} &    3.03\E{- 7} &    4.25\E{- 5} &    8.62\E{- 2} &    4.68\E{- 1} &    1.06\EE     &    1.92\EE     &    3.17\EE     &    4.66\EE     \\
\I{ 66}{Zn} &    8.44\E{-11} &    2.08\E{-10} &    2.46\E{-10} &    4.86\E{-10} &    4.47\E{-10} &    3.23\E{-10} &    1.53\E{-10} &    9.37\E{- 5} &    2.27\E{- 1} &    1.30\EE     &    3.07\EE     &    5.80\EE     &    1.02\E{  1} &    1.68\E{  1} \\
\I{ 67}{Zn} &    1.10\E{-10} &    6.82\E{-11} &    6.29\E{-11} &    6.69\E{-11} &    4.20\E{-11} &    6.25\E{-11} &    7.97\E{-12} &    4.41\E{- 7} &    1.23\E{- 3} &    7.38\E{- 3} &    1.79\E{- 2} &    3.45\E{- 2} &    6.26\E{- 2} &    1.06\E{- 1} \\
\I{ 68}{Zn} &    1.36\E{-10} &    6.52\E{-11} &    5.93\E{-11} &    6.23\E{-11} &    3.65\E{-11} &    5.97\E{-11} &    7.05\E{-12} &    2.31\E{- 8} &    7.78\E{- 5} &    5.15\E{- 4} &    1.31\E{- 3} &    2.65\E{- 3} &    5.08\E{- 3} &    9.13\E{- 3} \\
\I{ 69}{Ga} &    1.93\E{-10} &    1.51\E{-10} &    1.30\E{-10} &    1.33\E{-10} &    6.79\E{-11} &    1.14\E{-10} &    6.92\E{-12} &    6.86\E{- 9} &    8.21\E{- 5} &    6.03\E{- 4} &    1.56\E{- 3} &    3.10\E{- 3} &    5.56\E{- 3} &    8.89\E{- 3} \\
\I{ 71}{Ga} &    2.65\E{-10} &    1.19\E{-10} &    1.01\E{-10} &    1.04\E{-10} &    5.13\E{-11} &    8.72\E{-11} &    4.60\E{-12} &    3.12\E{- 9} &    8.49\E{- 5} &    6.93\E{- 4} &    1.85\E{- 3} &    3.75\E{- 3} &    6.91\E{- 3} &    1.15\E{- 2} \\
\I{ 70}{Ge} &    2.36\E{-10} &    4.40\E{-10} &    4.41\E{-10} &    4.67\E{-10} &    2.49\E{-10} &    4.37\E{-10} &    2.62\E{-11} &    9.36\E{- 7} &    2.04\E{- 2} &    1.62\E{- 1} &    4.26\E{- 1} &    8.56\E{- 1} &    1.57\EE     &    2.61\EE     \\
\I{ 72}{Ge} &    1.82\E{-10} &    1.15\E{-10} &    9.23\E{-11} &    9.18\E{-11} &    4.31\E{-11} &    7.24\E{-11} &    3.61\E{-12} &    5.64\E{-10} &    2.96\E{- 5} &    2.74\E{- 4} &    7.87\E{- 4} &    1.69\E{- 3} &    3.37\E{- 3} &    6.23\E{- 3} \\
\I{ 73}{Ge} &    1.71\E{-10} &    9.87\E{-11} &    8.91\E{-11} &    9.28\E{-11} &    4.61\E{-11} &    8.08\E{-11} &    4.10\E{-12} &    8.80\E{-12} &    9.81\E{- 7} &    9.82\E{- 6} &    2.80\E{- 5} &    5.85\E{- 5} &    1.07\E{- 4} &    1.69\E{- 4} \\
\I{ 75}{As} &    1.26\E{-10} &    1.47\E{-10} &    1.25\E{-10} &    1.28\E{-10} &    6.22\E{-11} &    1.10\E{-10} &    5.21\E{-12} &    6.30\E{-12} &    1.39\E{- 6} &    1.60\E{- 5} &    4.78\E{- 5} &    1.02\E{- 4} &    1.90\E{- 4} &    3.13\E{- 4} \\
\I{ 74}{Se} &    4.10\E{-12} &    3.25\E{- 9} &    4.09\E{- 9} &    4.65\E{- 9} &    2.55\E{- 9} &    4.98\E{- 9} &    2.54\E{-10} &    6.09\E{- 9} &    2.01\E{- 3} &    2.24\E{- 2} &    6.64\E{- 2} &    1.41\E{- 1} &    2.62\E{- 1} &    4.30\E{- 1} \\
\I{ 76}{Se} &    3.39\E{-10} &    6.41\E{-10} &    5.39\E{-10} &    5.30\E{-10} &    2.29\E{-10} &    3.96\E{-10} &    1.61\E{-11} &    1.06\E{-11} &    1.75\E{- 6} &    2.32\E{- 5} &    7.50\E{- 5} &    1.69\E{- 4} &    3.38\E{- 4} &    6.06\E{- 4} \\
\I{ 77}{Se} &    1.71\E{-10} &    8.40\E{-11} &    7.63\E{-11} &    8.00\E{-11} &    3.91\E{-11} &    7.12\E{-11} &    3.26\E{-12} &    1.67\E{-12} &    6.80\E{- 9} &    9.68\E{- 8} &    3.23\E{- 7} &    7.39\E{- 7} &    1.47\E{- 6} &    2.56\E{- 6} \\
\I{ 82}{Se} &    2.43\E{- 9} &    1.19\E{- 9} &    1.01\E{- 9} &    1.02\E{- 9} &    4.39\E{-10} &    8.51\E{-10} &    2.72\E{-11} &    1.20\E{-11} &    6.29\E{-10} &    5.66\E{-10} &    1.32\E{-12} &    9.94\E{-11} &    2.79\E{-11} &    3.73\E{-12} \\
\I{ 79}{Br} &    1.75\E{-10} &    1.43\E{-10} &    1.35\E{-10} &    1.39\E{-10} &    6.52\E{-11} &    1.19\E{-10} &    4.94\E{-12} &    2.41\E{-12} &    3.51\E{- 9} &    5.24\E{- 8} &    1.80\E{- 7} &    4.17\E{- 7} &    8.50\E{- 7} &    1.53\E{- 6} \\
\I{ 81}{Br} &    6.16\E{-10} &    6.16\E{-10} &    5.58\E{-10} &    5.72\E{-10} &    2.55\E{-10} &    4.83\E{-10} &    1.71\E{-11} &    7.83\E{-12} &    3.66\E{-10} &    5.31\E{-10} &    7.96\E{-10} &    1.98\E{- 9} &    4.01\E{- 9} &    7.28\E{- 9} \\
\I{ 78}{Kr} &    8.50\E{-25} &    8.00\E{-10} &    1.92\E{- 9} &    2.75\E{- 9} &    1.87\E{- 9} &    4.66\E{- 9} &    2.18\E{-10} &    1.25\E{-10} &    6.39\E{- 9} &    1.54\E{- 8} &    3.25\E{- 8} &    7.61\E{- 8} &    1.53\E{- 7} &    2.77\E{- 7} \\
\I{ 80}{Kr} &    4.42\E{-19} &    1.58\E{- 9} &    2.89\E{- 9} &    3.51\E{- 9} &    1.91\E{- 9} &    4.24\E{- 9} &    1.63\E{-10} &    8.18\E{-11} &    4.85\E{- 8} &    7.71\E{- 7} &    2.76\E{- 6} &    6.58\E{- 6} &    1.37\E{- 5} &    2.52\E{- 5} \\
\I{ 83}{Kr} &    6.48\E{-10} &    4.23\E{-10} &    3.82\E{-10} &    3.94\E{-10} &    1.64\E{-10} &    3.47\E{-10} &    9.38\E{-12} &    4.08\E{-12} &    2.91\E{-10} &    6.14\E{-10} &    1.36\E{- 9} &    3.38\E{- 9} &    7.05\E{- 9} &    1.31\E{- 8} \\
\I{ 84}{Sr} &    6.77\E{-22} &    1.54\E{-10} &    6.12\E{-10} &    1.03\E{- 9} &    8.28\E{-10} &    2.80\E{- 9} &    1.11\E{-10} &    6.58\E{-11} &    5.67\E{- 9} &    1.94\E{- 8} &    5.47\E{- 8} &    1.40\E{- 7} &    3.03\E{- 7} &    5.83\E{- 7} \\
\hline
\PPI&\PPE&\PPE&\PPE&\PPE&\PPE&\PPE&\PPE&\PPE&\PPE&\PPE&\PPE&\PPE&\PPE&\PPE \\
\end{tabular}
}\vspace{-1.5\baselineskip}
\begin{flushright}\textsc{(end of production factor table)}\end{flushright}
\end{table}

\end{landscape}

}

\ifthenelse{\boolean{emul}}{ }{

\clearpage
\onecolumn

\ifthenelse{\boolean{\IncludeFigures}}{
\renewcommand{\figcaption}[2][]{
\clearpage
\begin{figure}
\epsscale{0.6}
\plotone{#1}
\caption{#2}
\end{figure}
}}{}

\figcaption[\FigyeFile]{\Figye}
\figcaption[\FigMMFile]{\FigMM}
\figcaption[\FigpfFile]{\Figpf}
\figcaption[\FigpfmFile]{\Figpfm}
\figcaption[\FigpfMFile]{\FigpfM}
}

\end{document}